\documentclass{article}

\usepackage{arxiv}

\usepackage[utf8]{inputenc} 
\usepackage[T1]{fontenc}    
\usepackage{hyperref}       
\usepackage{url}            
\usepackage{booktabs}       
\usepackage{amsfonts}       
\usepackage{nicefrac}       
\usepackage{microtype}      
\usepackage{lipsum}		
\usepackage{graphicx}
\usepackage{natbib}
\usepackage{doi}

\usepackage{float}
\usepackage{soul}
\usepackage{xargs}                      
\usepackage[pdftex,dvipsnames]{xcolor}  
\usepackage[colorinlistoftodos,prependcaption,textsize=footnotesize]{todonotes}
\newcommandx{\todoAndres}[2][1=]{\todo[linecolor=blue,backgroundcolor=blue!25,bordercolor=blue,#1]{#2}}

\usepackage{amsmath,amssymb}
\usepackage[linesnumbered,boxed]{algorithm2e}
\usepackage{hyperref}
\usepackage[capitalize,nameinlink]{cleveref}

\usepackage{subcaption}
\usepackage{mwe}

\newcommand{\mlold}[1]{{\color{blue}{#1}}}
\newcommand{\andres}[1]{{\color{purple}{#1}}}
\renewcommand{\mlold}[1]{{#1}}

\newcommand{\ml}[1]{{\color{blue}{#1}}}
\renewcommand{\ml}[1]{{#1}}
\renewcommand{\andres}[1]{{#1}}

\usepackage{array}
\usepackage{tabularx}
\usepackage{wrapfig}
\usepackage{picins}

\usepackage{multirow,booktabs}
\usepackage{makecell}
\usepackage{comment}

\title{A Lightweight Semi-Centralized Strategy for the Massive Parallelization of Branching Algorithms}

\date{April 29, 2023}	

\author{ \href{https://orcid.org/0000-0001-9813-0097}{\includegraphics[scale=0.06]{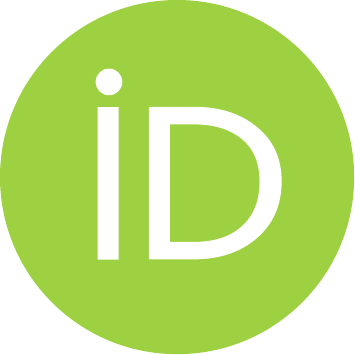}\hspace{1mm}Andres ~Pastrana-Cruz}\\
	Département d'informatique\\
        Université de Sherbrooke\\
	2500 Boulevard de l'Université \\
        Sherbrooke,J1K 2R1, QC, \\
        Canada\\
	\texttt{Andres.Pastrana@usherbrooke.ca} \\
	\And
	\href{https://orcid.org/0000-0002-5305-7372}{\includegraphics[scale=0.06]{orcid.pdf}\hspace{1mm}Manuel ~Lafond}\thanks{Corresponding author} \\
	Département d'informatique\\
        Université de Sherbrooke\\
	2500 Boulevard de l'Université \\
        Sherbrooke,J1K 2R1, QC, \\
        Canada\\
	\texttt{Manuel.Lafond@usherbrooke.ca} \\
}

\renewcommand{\shorttitle}{A lightweight semi-centralized strategy for the massive parallelization of branching algorithms}

\hypersetup{
pdftitle={A template for the arxiv style},
pdfsubject={q-bio.NC, q-bio.QM},
pdfauthor={Andres ~Pastrana-Cruz, Manuel ~Lafond},
pdfkeywords={Load balancing, Vertex Cover, Parallel algorithms, Scalable parallelism, Branching algorithms},
}

\begin{document}
\maketitle

\begin{abstract}
	
Several NP-hard problems are solved exactly using exponential-time branching strategies, whether it be branch-and-bound algorithms, or bounded search trees in fixed-parameter algorithms. 
\mlold{The number of tractable instances that can be handled by sequential algorithms is usually small, whereas}
massive parallelization has been shown to significantly increase the \mlold{space} of instances that can be solved exactly.  However, previous centralized approaches require too much communication to be efficient, whereas decentralized approaches are more efficient but have difficulty keeping track of the global state of the exploration.  

In this work, we propose to revisit the centralized paradigm while avoiding previous bottlenecks.  In our strategy, the center has lightweight responsibilities, requires only a few bits for every communication, but is still able to keep track of the progress of every worker.  In particular, the center never holds any task but is able to guarantee that a process with no work always receives the highest priority task globally.

Our strategy was implemented in a generic C++ library called GemPBA, which allows a programmer to convert a sequential branching algorithm into a parallel version by changing only a few lines of code.  An experimental case study on the vertex cover problem \mlold{demonstrates that some of the toughest instances from the DIMACS challenge graphs that would take months to solve sequentially can be handled within two hours with our approach}.
\end{abstract}


\section{Introduction}
\label{sec:introduction}

Several scientific disciplines require solving NP-hard problems for which no polynomial-time algorithm is believed to exist.
This includes, for instance, clustering proteins in biological networks~\cite{ideker2008protein}, maximizing influence in a social network~\cite{kempe2003maximizing}, 
or optimizing weights in neural networks~\cite{judd1990neural}.
Such NP-hard problems are usually handled by fast heuristics or approximation algorithms when the running times are crucial.
However, the recent ease of access to high-performance architectures, combined with novel algorithmic techniques, have allowed researchers to aim for exact algorithms in reasonable times, even if above polynomial.  
These are usually exponential in the input size, 
and recent research has focused on making the algorithmic complexity tolerable for some practical purposes~\cite{fomin2013exact} (for instance, achieving a complexity of $O^*(1.23^n)$ for maximum independent set~\cite{fomin2009measure}, where $O^*$ suppresses polynomial factors). 
Branch-and-bound algorithms, which brute-force every possible solution but skip those that cannot do better than the current optimal, have also been studied extensively~\cite{clausen1999branch,morrison2016branch}.
Another recent research trend consists of fixed-parameter tractability (FPT), which aims to design algorithms that are exponential, but only with 
respect to a parameter that is expected to be small~\cite{niedermeier2006invitation,downey2013fundamentals,cygan2015parameterized}.  We also refer the reader to~\cite{woeginger2003exact} for an excellent survey of exact algorithms for NP-hard problems.

In this paper, we focus on improving the scalability of such approaches with massive parallelization.
A well-known technique for both exponential and FPT algorithms is recursive \emph{search tree exploration}.  In a nutshell, when given an instance $I$ to solve, search tree algorithms generate a few sub-instances of $I$
in a way that at least one of them leads to an optimal solution.  The algorithm then explores each sub-instance recursively until a solution is found, or until 
the whole space has been searched, depending on the algorithm.  This forms a recursion tree in which \emph{nodes} correspond to function calls and 
\emph{children} correspond to its recursive calls.

There is an extensive literature on the problem of parallelizing search tree algorithms.
If $p$ processes are available, the most straightforward strategy is to assign each node at depth $\log p$ its own process, and see which one finds a solution~\cite{bokhari1979,cheetham2003solving}.
This works in an idealized setting where each process is assigned a tree of about the same height, but most search trees are unbalanced and 
some processes will finish before others.
In this case, they can and should be reassigned to other subtrees of the search tree, leading to the problem of dynamic load-balancing.

It is not obvious how to perform this optimally, since
massive parallelization of search trees introduces two problems: how to distribute search trees to processes, and how to minimize communication.  Unsurprisingly, there is an inherent tradeoff to choose from between communication overhead and search efficiency, as more communication allows assigning free processes to the most important or most promising task globally available.  A \emph{centralized} strategy 
was developed by Abu-Khzam et al.~\cite{Abu-Khzam2006} with this idea in mind, where a center would maintain a queue of available tasks and nodes, and thus could always make optimal assignment choices.
As argued later by Abu-Khzam et al.~\cite{abu2015scalable}, communication overhead is not worth the gained efficiency, especially since tasks may require sending lots of information.  An opposite \emph{decentralized} strategy was therefore developed.  The idea is to arrange the cores into a virtual hierarchy and let cores only accept tasks from their superior.  This optimizes communication but no core has an idea of the global situation and, as we argue in this paper, this leads to suboptimal task assignment and exploration.
We also refer the reader to~\cite{shu1989,saletore1990consistent,kale1992prioritization,sinha1993load,abu2007buffered,sun2011adaptive,Weerapurage2011} for further ideas that have focused on how to explore tasks efficiently in parallel.  These works all aim to describe search tree distribution paradigms that are applicable to any branch-and-bound or branching FPT algorithm, of which the framework of~\cite{abu2015scalable} is the latest, to our knowledge.

However, it is important to note that parallel search tree algorithms are still being developed actively, usually for specific problems or architectures. 
\mlold{Notably, Archibald et al. recently developed YewPar~\cite{archibald2018replicable,archibald2020yewpar,macgregor2022generic}, a generic framework created independently from our work but with similar goals.  The framework implements several known scheduling strategies for search algorithms and allows the integration of custom schedulers (although they do not propose semi-centralized strategies).} 
Some other works implement standard load balancing approaches (work pool, work-stealing) with the aim of studying the impact of hardware or architectural designs, for instance multi-core versus many-core, vectorization schemes~\cite{melab2018multi}, or access to an elasticity controller~\cite{kehrer2020equilibrium}.
In~\cite{rauchecker2019using,leoncini2019parallel,soto2020solving}, the authors propose parallelism strategies tailored for various job scheduling problems and in~\cite{smirnov2018domain,fallah2021parallel}, parallel branch and bound strategies are proposed for mixed integer linear programs.
Also note that~\cite{melab2018multi,rauchecker2019using} make use of a centralized strategy in their parallel implementation, which are similar to the fully centralized idea developed in~\cite{Abu-Khzam2006}.

\noindent 
\textbf{Our contributions.}
In this paper, we propose a novel \emph{semi-centralized} load-balancing strategy that takes advantage of both approaches and offers a balance in the communication versus exploration tradeoff. 
The semi-centralized name comes from the fact that not all important communication needs to pass through the center.  The main idea is to make use of a central process, but in an extremely lightweight fashion, in the sense that (1) communication with the center is asynchronous; (2) each message is small as it only requires sending a single integer, and (3) communication is limited since messages are only exchanged with the center when a worker is finished or when a better solution is found.  The center is relieved from the heavy responsibility of maintaining a task queue, and instead is only present to maintain the status of working processes and dynamically decide which processes should exchange tasks.  The heavy task communication is only performed between working processes that need to share information, and only when it is necessary to do so.    In particular, an idle working process only needs to request work once, in contrast with previous solutions where such requests could fail and require multiple communication rounds.  We also propose a strategy that allows each worker to maintain the hierarchy of its highest priority search tasks which is applicable to search tree algorithm with any branching factor, even if it is heterogeneous across the search tree. 
Moreover, our approach allows a \emph{process-thread} hybrid implementation.  That is, a subtree assigned to a process can be partitioned into further subtrees, each assigned to a different thread.  
Our strategy is implemented in a generic, open-source C++ library called {\sf GemPBA}.
The library uses Message Passing Interface (MPI), is user-centric, and a programmer is able to parallelize any existing sequential search tree function by changing a few lines of code.

We use the traditional vertex-cover problem as a case study.  
We demonstrate that even with the simplest branching implementation for vertex-cover, our library achieves close to linear speedup and can solve some of the toughest instances of the DIMACS challenge graphs~\cite{bader201110th}.  \mlold{We compare our approach with a fully centralized strategy, and show that it cannot surpass the times achieved by the semi-centralized strategy.  Moreover, we demonstrate that the efficiency of the centralized strategy is heavily dependent on how tasks are serialized, and that the semi-centralized approach is more robust to this aspect.}

\section{Preliminary notions}\label{sec:prelim}

In this section, we first explain the branch-and-bound and fixed-parameter search tree algorithms at a high level.  
To motivate the need for novel ideas in parallelizing these algorithms, we then discuss the main load-balancing strategies that have been applied to branching algorithms in the literature, along with their advantages and disadvantages.

\subsection{Search tree algorithms}

In essence, all branching algorithms have a similar structure.
Given an instance $I$, we first verify whether $I$ is a solution to our problem, which corresponds to a terminal case.
Otherwise, we generate a set of (usually smaller) instances $I_1, \ldots, I_k$ from $I$ in a way that at least one $I_j$ can lead to a solution.  We then explore each $I_j$ recursively. 

\begin{algorithm}
\DontPrintSemicolon
\SetKwProg{Fn}{function}{}{}
\Fn{searchTree($I$)}
  {
    \uIf{$I$ cannot lead to a solution better than $best$}
    {
        return\;
    }
    \uIf{$I$ is a solution}
     {
        \uIf{$I$ is better than $best$}
        {
            $best = I$\;
        }
        return \;
     }
     
     $I_1, I_2, \ldots, I_k = $ sub-instances of $I$\;
     \For{$j = 1..k$}
     {
        searchTree($I_j$)\;
     }
  }     
  \caption{Structure of a sequential search tree algorithm.}
  \label{alg:seqalgo}
\end{algorithm}

\vspace{3mm}

We distinguish branch-and-bound and fixed-parameter algorithms, which we briefly describe since our methodology applies to both.
In the well-known \emph{Branch-and-Bound} (B\&B) paradigm, we must optimize some value and the best solution found so far is stored globally.  
When reaching a terminal case, we check whether the solution is better than the best, and if so we update it.
More importantly, whenever an instance is guaranteed to lead to a worse solution than the best, we stop the recursion.  The \emph{branching factor} is the maximum number of recursive calls the algorithm makes.
Algorithm~\ref{alg:seqalgo} presents the general structure of this type of algorithm.
In the \emph{fixed-parameter tractability} (FPT) paradigm, we instead have a decision problem that asks whether there exists a solution of size $k$.
If $I$ cannot lead to such a solution, we can return.  If $I$ is such a solution, we can return ``yes'' and stop all exploration (contrasting with branch-and-bound, which keep exploring).
Notably, fixed-parameter algorithms are known for \emph{kernelization}, which describe rules to reduce the $I$ instance to a smaller size (see e.g.~\cite{flum2006parameterized,cygan2015parameterized}).
Several parallelism ideas have been proposed for FPT algorithms~\cite{cheetham2003solving,abu2015scalable,bannach2019towards}.

As a concrete example of a search tree algorithm, consider the \emph{vertex cover} problem.  In the optimization version, we receive a graph $G = (V, E)$ and must find a subset $X \subseteq V$ of minimum size that touches every edge, i.e. for all $uv \in E$, $u \in X$ or $v \in X$.
A simple branching strategy goes as follows.  Each recursion receives a partial solution $S \subseteq V$, with $S = \emptyset$ at the initial call.
We choose $uv \in E$ not covered by $S$ and observe
that we can either 1) add $u$ to $S$; 2) not add $u$ to $S$.
In the second case, all the neighbours $N(u)$ of $u$ must be added to $S$ to cover its incident edges.  We thus recursively branch into two subinstances: $G - \{u\}$ with partial solution $S \cup \{u\}$, and $G - N(u)$ with partial solution $S \cup N(u)$ ($G - X$ is the graph obtained by removing the $X$ vertices).  The recursive calls then check whether these partial solutions have more vertices than the current best solution, and if so do not explore it.  In terms of FPT, the algorithm is the same, except that we stop exploring if the current partial solution has more vertices than $k$, the parameter.
Vertex cover is a standard problem that is used for several experimental benchmarking tools~\cite{Abu-Khzam2006,abu2015scalable,wang2019exact,abu2018accelerating}.

\subsection{Previous search tree parallelization strategies}

Several approaches have been proposed to parallelize branching algorithms.  We present the main categories that we have identified, with an emphasis on full decentralization, since it has been reported to be able to solve the most difficult vertex cover instances.

\paragraph{Equitable parallelization}
Assume that $p$ processors are available and that the branching factor is $r$.
In~\cite{bokhari1979}, the authors propose to execute the algorithm sequentially until a depth of $\log_r p$ is reached.  This defines a tree with $p$ leaves corresponding to $p$ instances, at which point each processor is assigned a distinct instance.
This distributes the search tree across processes ``equally'', but an obvious disadvantage of this approach is that, once a process has finished exploring the search tree of its assigned instance, it is not recycled to help exploring other subtrees. 
This strategy was also used in~\cite{cheetham2003solving} to solve FPT problems (along with several other strategies, including the usage of free processors for faster kernelization).  

\paragraph{Greedy load balancing}
In~\cite{shu1989}, the authors apply process recycling to the search tree exploration.  
When a recursive call needs to branch into a new subtree $T$, if some process $p_i$ is available, then it assigns $p_i$ to $T$, and otherwise explore sequentially.  We must assume that each free $p_i$ broadcasts its availability to the others.  We call this the greedy approach since a process assigns its most recent task to the most recently freed process as soon as possible, regardless of the current state of the search tree.  
The advantage is that processes are constantly participating in some tree exploration.  However, this strategy tends to assign processes \emph{vertically}.  That is, once a process $p_i$ is assigned to a subtree it will start digging deeper and deeper into it.  When another process $p_j$ gets freed, it will be assigned to the current location of $p_i$, which is likely to be deep in its recursion.
This tends to bias the exploration to similar parts of the search tree.  Moreover, free processes get assigned small search trees, leading them to finish quickly and broadcast information more often.
In~\cite{Weerapurage2011}, a hybrid strategy is proposed, where processes are first distributed in an equitable manner, and then reassigned greedily.

\paragraph{Fully centralized approach}

In~\cite{Abu-Khzam2006}, Abu-Khzam et al. have developed a strategy where a central process is responsible for receiving tasks from the other processes.  The center maintains a queue of tasks of bounded size.  When another process requests a task, the center can choose which one to send according to some priority function.  Examples of priority include the task with the largest subtree to explore, or the task with the most promising solution so far.
This is an important advantage, since such a priority scheme avoids the vertical-exploration problem mentioned above.
The main drawback is the large communication overhead for constant requests to center.
Also, exponential algorithms tend to saturates the task queue very quickly. The ability for the center to choose tasks with priority is therefore hindered by the fact that most tasks never make it into the queue, and in the end, there is little control over task priority.  
Let us also mention that in~\cite{Weerapurage2011}, the authors developed a centralized scheduler-based strategy specifically for the FPT vertex cover problem.  

\paragraph{Fully decentralized approach}

To address the problems of full centralization, 
Abu-Khzam et al. explored the other extreme by devising a fully decentralized approach~\cite{abu2015scalable}.  
In this strategy, the available cores are organized into a tree  (which should be distinguished from the search tree).
In this topology, the parent of core number $r$ is $r - 2^{\lfloor \log(r) \rfloor}$. 
Initially, the root of the core-tree is assigned the full instance.  Then, each core requests a task to its parent in the core-tree.  
The degree distribution of the core tree is heaviliy skewed towards nodes near the root.  The root of the core-tree has around $\log c$ children, while the majority of nodes have $0$ or $1$ children, where $c$ is the number of cores.  This topology is chosen because it initially assigns cores as in the equitable strategy.

When a core is finished with its instance, it asks its parent for a new task asynchronously, and if none is available, the request fails and it switches to a new parent.  This strategy requires no synchronization, since there is no way a process can receive multiple tasks.  Another important aspect of this strategy is that when a core $r$ has a pending request from a child core $q$, $r$ chooses to give $q$ the sub-instance that is the highest in its search tree.  An indexing of tasks is proposed to maintain the highest priority, which is the highest unexplored node in the search tree.  The vertical-exploration problem is therefore avoided, although each core has its own priority instead of having a center that maintains global priority.
Therefore, a worker receives the most urgent task from its parent, not necessarily the most globally urgent task.
Moreover, because the core-tree is imbalanced, some cores are more likely to receive help than others.  For instance, a subtree assigned to a leaf of the core-tree will never receive help (unless a lucky parent reassignment occurs), whereas $\log c$ cores are available to assist the subtree assigned to core 0.
This inherent bias is difficult to circumvent in a fully decentralized setting, which motivates our new strategy.
Finally, let us also mention that work requests can fail, which occurs when a parent has no work for a child.  This forces the child to send another request to another core and, as the experiments in~\cite{abu2015scalable} show, this can have a significant performance impact on difficult graphs.

\section{A semi-centralized load-balancing strategy}

As we have discussed, centralized and decentralized strategies each have their own pros and cons.  Here, we propose a novel strategy that is in-between.  
We do make use of a central process, but its responsibilities are reduced to a minimum, as well as its communication and memory requirements.  In particular, tasks do not go through the center, as it only stores the smallest amount of information required to know which processes should exchange tasks.
The center is designed with three goals in mind:

\begin{enumerate}

    \item 
    The center must never become overloaded, and its memory usage should be independent of the number of ongoing or pending tasks;
    
    \item 
    Communication and synchronization should be minimized.  In particular, a process should never be waiting for a reply from another process, unless it has no task to work on.  Also, failed work requests should be minimized or non-existent.
    
    \item 
    When a process is available, it should be possible to assign it to the task with highest global priority.  Moreover, the current best solution found globally should always be available to every process.

\end{enumerate}

Let us also mention that our strategy is designed to achieve the ease of use illustrated in Algorithm~\ref{alg:schedalgo}. 
This shows the same algorithmic structure as Algorithm~\ref{alg:seqalgo}, but in which all parallelism is handled by our GemPBA library.  
\mlold{That is, when an instance $I$ is split into a set of child instances $I_1, \ldots, I_k$ to explore, the latter are registered to a scheduler instead of being explored immediately.  The scheduler then tracks these sub-instances and launches corresponding tasks only when appropriate.  The scheduler can also prioritize these sub-instances and even remove some of them from its queue if they cannot lead to a better solution than the current one.}
Such a delegation scheme allows almost any sequential branching algorithm to be parallelized with minor modifications.

\begin{algorithm}
\DontPrintSemicolon
\SetKwProg{Fn}{function}{}{}
\Fn{searchTree($I$)}
  {
    $best$ = GemPBA.getBestSolution()\;
    \uIf{$I$ cannot lead to a solution better than $best$}
    {
        return\;
    }
     \uIf{$I$ is a solution}
     {
        GemPBA.handleSolution($I$)\;
        return \; \label{line:returnaftersol}
     }
     
     $I_1, I_2, \ldots, I_k = $ sub-instances of $I$\;
     GemPBA.\mlold{registerChildInstances}($I_1, \ldots, I_k$, $parent = I$)\;
     \For{$j = 1..k$}
     {
        GemPBA.search($I_j$) \tcp*{Library handles parallelism}
     }
  }     
  \caption{Structure of a parallelized search tree algorithm.  }
  \label{alg:schedalgo}
\end{algorithm}
\vspace{3mm}

\mlold{Let us note that in some cases, it might be advantageous to remove the \emph{return} statement on line~\ref{line:returnaftersol}, for instance if further exploration of the sub-instances of a solution $I$ could lead to improved bounds for other search nodes.  It can thus be seen as optional.}
Our load-balancing is also required to be compatible with a combination of multithreaded and multiprocess exploration, so that processes could use thread for their exploration without affecting the above goals.

\subsection{Center and worker responsibilities}

In addition to handling startup and termination, the center is only responsible for:

\begin{itemize}
    
    \item Maintaining the list of processes available to receive new tasks, and possibly other metadata on each process;
    \item Determining which working process should send a task to which available node, without conflict;
    \item Maintaining the value of the best global solution so far \mlold{and the worker that currently holds this best solution}.
\end{itemize}

One can see that in terms of storage, center only needs to remember a simple array of process statuses, and a numerical value for the best solution.  \mlold{The center also knows which worker found this solution, and center only fetches the optimal solution when the whole exploration has finished.} 
The metadata is optional and is intended to store the priority (an integer) of the most urgent task in each process.  
This metadata can let center determine, when a process $p_i$ is available, which process $p_j$ has the heaviest task that can be sent to $p_i$ (without center having to actually store the task).
In that manner, center can let $p_i$ know that it should send it to $p_j$.  Other assignments are possible if no metadata is present, for instance by choosing $p_j$ randomly.\\

\noindent
The workers are responsible for:

\begin{itemize}
    \item Exploring the search tree of a given instance;
    \item Communicating to center to request work;
    \item \mlold{Maintaining the best solution found locally so far, and informing center of the value of this solution when appropriate};
    \item Sending their heaviest pending task to processes assigned by center.
\end{itemize}

Note that we aim for work requests to never fail, i.e. once a process sends an availability message, it will always receive a task without having to ask again.  
This contrasts with the decentralized setting, where work requests may fail when the process that we send a request to has no available tasks, in which case we must try another process.  In fact,\cite{abu2015scalable} reported that the number of failed requests has been shown to be significantly high.  
Our failure-free guarantees are achieved by ensuring that center assigns a worker to the available process that is guaranteed to send it work eventually.
Also note that maintaining the heaviest pending tasks is not trivial.  Later in this section, we propose a strategy that stores the state of the pending search tree nodes to achieve this, while only requiring an amount of memory that grows linearly.

\begin{figure}[ht]
    \centering
    \includegraphics[width=0.7\textwidth]{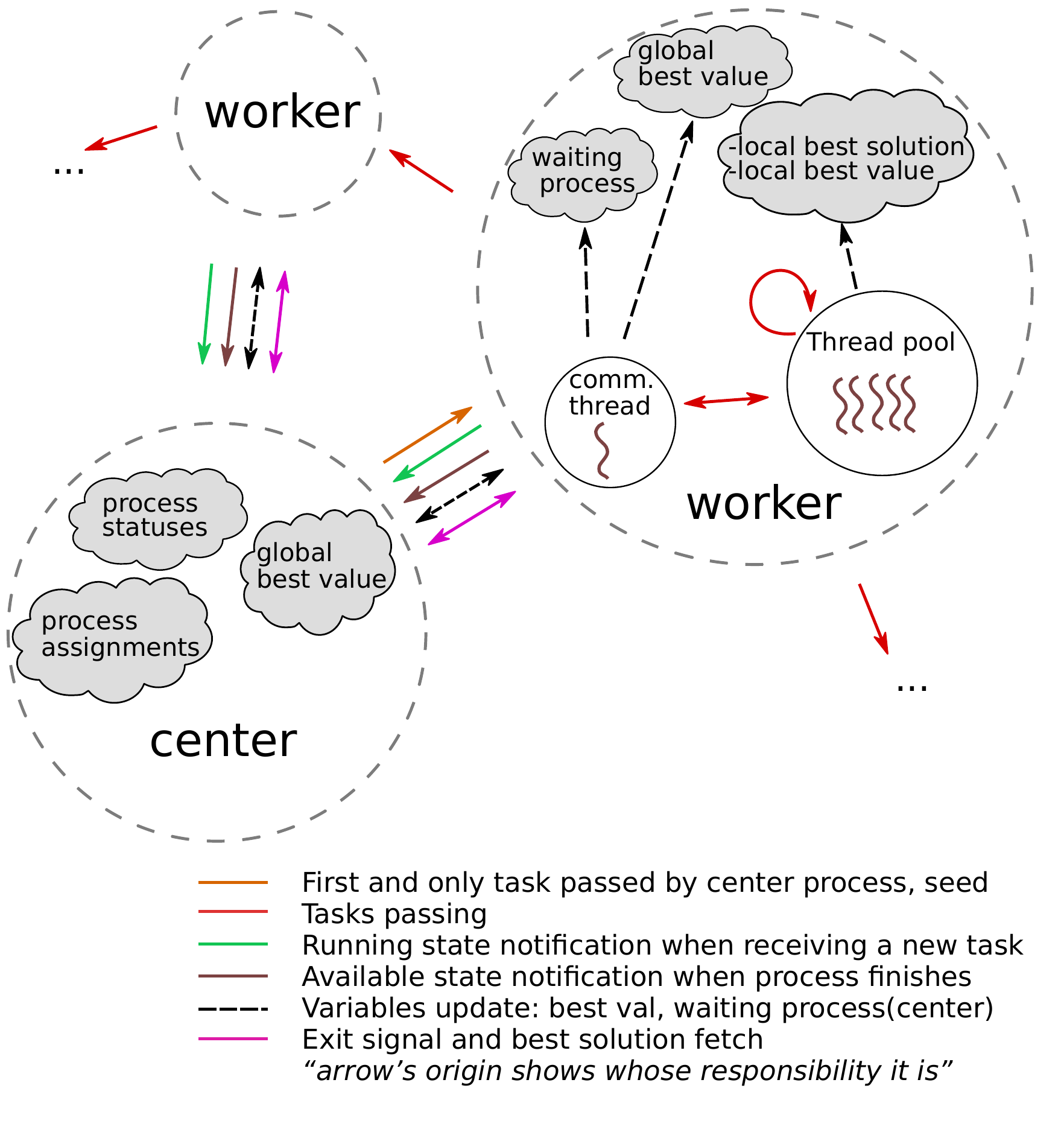}
    \caption{Communication topology.  The ``seed'' refers to the original instance that  is sent to the first working process to initiate the exploration.}
    \label{fig:fig1}
\end{figure}

\cref{fig:fig1} illustrates the relationship between processes.

\subsection{Center implementation}

Given its lightweight set of responsibilities, the center is relatively simple to implement.  The high level ideas are shown in Algorithm~\ref{alg:runcenter}.  The reader should bear in mind that whenever a message is sent, there is no need to wait for a reply or a confirmation --- everything occurs asynchronously.  
The center is in a listening loop and reacts to worker messages.  To manage the optimal value, it stores a single numerical value $bestValSoFar$ to remember the global optimal and, when a worker thinks it has found a better value, it receives a $bestval\_update$ request.  Since several such updates can be received in a short time span, center needs to verify this claim, but this is an easy check.  If the best value indeed changes, center broadcasts this to all processes in a non-blocking fashion, so that workers can update their local best value when they have time.

\begin{algorithm}
\DontPrintSemicolon
\SetKwProg{Fn}{function}{}{}
\Fn{centerLoop()}
  {
    Recv($tag, source, data$)\;
    \uIf{$tag == $ ``bestval\_update'' and $data < bestValSoFar$}
    {
            $bestValSoFar = data$\;
            Async\_Broadcast(tag = $bestval\_update$, data = $data$)\;
    }
    \uElseIf{$tag == $ ``available''}
    {
        $w = getNextWorkingNode()$\;
        \uIf{$w$}
        {
            Async\_Send(dest = $w$, tag = $send\_work$, data = $source$)\;
            $status[source] = ASSIGNED$\;
        }
        \uElse 
        {
            $status[source] = AVAILABLE$\;
        }
    }
    \uElseIf{$tag == $ ``started\_running''}
    {
        $status[source] = RUNNING$\;
        \uIf{$\exists w$ such that $status[w] == $ AVAILABLE}
        {
            Async\_Send(dest = $source$, tag = $send\_work$, data = $w$)\;
            $status[w] = ASSIGNED$\;
        }
    }
    \uElseIf{$tag == $ ``metadata''}
    {
        $updateMetaData(source, data)$\;
    }
  }

  \uIf{$status[p] == AVAILABLE$ for every process $p$}
  {
      tryTermination() \tcp{see Section~\ref{sect:worker}}
  }

  \caption{Pseudocode of the center loop (assuming a minimization problem)}
  \label{alg:runcenter}
\end{algorithm}

The center can also receive messages when a worker changes state.  An array of states is maintained, with one entry per process.  When a worker $r$ has finished exploring its subtree, center receives an $available$ message.  At this point, center chooses a worker $w$ with the $getNextWorkingNode$ function, and at this point $w$ should send its heaviest task to $r$.  
This choice can either be made randomly, or according to some priority function based on the metadata. 
More specifically, in our implementation, $getNextWorkingNode$ simply chooses $w$ uniformly at random among the processes that satisfy $status[w] = running$.  
Another possibility would be to expect the workers to regularly send metadata that contains the size of the largest unexplored \mlold{instance} (which is just one integer).  The center stores this metadata for each process, and another possible implementation of $getNextWorkingNode$ would be to pick the running process with the highest value in its metadata.  
The point here is that our framework allows several priority possibilities, but since our goal in this work is to validate the general methodology of the center, we reserve the study of the impact of the $getNextWorkingNode$ function for future works.

Importantly, the task to send does not go through center --- rather, center sends a non-blocking message to the chosen $w$ to let it know it must send a task to $r$, and then remembers that $r$ is waiting by putting its state to $assigned$.  This ensures that one and only one worker can now send a task to $r$, thereby avoiding conflict.  
Moreover, this assignment persists until $w$ 
does send work to $r$, ensuring that the work request from $r$ does not fail.  
At this point, it is the sole responsibility of $w$ to send a task to $r$, and the center does not need to check on the status on $w$ or $r$.  When $r$ receives work from $w$, it will update the center, although it is possible that $w$ has no work to send immediately.  In this case, $w$ will let center know that it has no job and $r$ will remain on \mlold{the waiting list of $w$}.  Center will then eventually ask some other process to send work to $w$, and the chain of task sending will resume.  \mlold{Note that before assigning $r$ to $w$, the center can follow the chain of assignments that starts at $r$ to ensure that it does not already lead to $w$, to ensure that no cyclic dependencies are introduced.}

When an idle worker $r$ receives a task, it lets center know by sending a $started\_running$ message.  \mlold{At this point, center may have a non-empty list of idle workers that are yet unassigned to another process (this could happen for brief moments, for instance when more than half of the workers complete their task almost at the same moment).  If this is the case, center assign to $r$ one of these unassigned workers.}

\subsection{Worker implementation}\label{sect:worker}

Workers can either send new information to the center or react to center messages.  
This is done by periodically calling the update functions $updateWorkerIPC$ and  \mlold{$updatePendingTasks$} displayed in~\cref{alg:updateworker}.
\andres{The function  $updateWorkerIPC$ is responsible for receiving messages from other processes, which can contain various information such as new bounds for the optimal solution, processes that are awaiting data, tasks to handle, and metadata (if applicable).}

The $updatePendingTasks$ function checks whether there are pending tasks and, if so, sends them to either process assigned by the center or other idle threads (we give priority to other processes).
The periodical calls to these functions could be implemented in two ways.  
They could be called at the start of every call to the $searchTree$ procedure in~\cref{alg:schedalgo}.
Another option is to dedicate a thread to updates and communications, which would call the functions in a loop.  
Although straightforward, the two options are worth mentioning because the former is not compatible with every interprocess communication library.  Indeed, in a multithreaded environment, the first solution allows \emph{any} thread to send remote messages, and \emph{openmpi} is  \andres{known to have} difficulty dealing with this~\cite{Gropp2006}.  
Fixing a looping thread for these tasks is, therefore, easier to implement --- the priority of this thread can be lowered or a small sleep can be added to make it use less CPU.
There is no thread interference between such a looping thread and the workers since the looping thread does not access the instances that are being explored.  The impact on the performance of this decision should therefore be limited.

\begin{algorithm}
\DontPrintSemicolon
\SetKwProg{Fn}{function}{}{}
\Fn{updateWorkerIPC()}
  {
    \uIf{$hasMessage()$}
    {
        $Recv(source, tag, data)$\;
        \uIf{$tag == bestval\_update$}
        {
            \uIf{$data < local\_bestval$}
            {
                $global\_bestval = local\_bestval = data$\;
            }
        }
        \uElseIf{$tag == send\_work$}
        {
            $addToWaitingProcesses(process = data)$\;
        }
        \uElseIf{$tag == work$}
        {
            //this can only be received when no task is running\;
            Async\_Send(dest = center, tag = $started\_running$)\;
            $sendTaskToNextThread(instance = data)$\;
        }
    }
    \uIf{$local\_bestval < global\_bestval$}
    {
        Async\_Send(dest = $center$, tag = $bestval\_update$, data = $local\_bestval$)\;
    }
    
    \uIf{$hasMetadata()$}
    {
        Async\_Send($dest = center$, tag = $metadata$, \quad data = $metadata$)\;
    }
  }
\;
\Fn{\mlold{updatePendingTasks()}}{
    \While{$hasPendingTasks()$ and $hasWaitingProcess()$}
    {
        \mlold{$dest =  getWaitingProcess()$\;}
        $sendHighestPriorityTask(dest)$\;
    }
     \While{$hasPendingTasks()$ and $thread\_pool.hasIdle()$}
    {
        dest = \mlold{getNextThread()}\;
        $sendHighestPriorityTask(dest)$\;
    }
}

  \caption{Pseudocode of worker update functions. }
  \label{alg:updateworker}
\end{algorithm}
\vspace{3mm}

In any case, let us emphasize that workers should never be in a blocking listening mode.  Instead, the update functions check whether center has left them a message in their receiving buffer and resume if not (which most libraries allow, for instance $MPI\_Iprobe$ using openmpi).
We assume that workers store a variable $local\_bestval$ for the optimal value found by its threads, and $global\_bestval$ for the optimal value seen by center.  
When center sends a better value than the local, both are updated (protected by a mutex, since threads can change the local \andres{variable}).  
Center can also ask to send work to another process $r$, in which case we add $r$ to our waiting list.  The update functions will eventually send work to those on the waiting list.  Note that in our implementation, the center ensures that each waiting list has at most one element (except at startup).

After checking for center messages, 
the worker can decide to send its local best value if it thinks it is better than center's (and the latter will verify this).
We also check whether a task can be sent to an idle thread if there is one, and metadata update can be sent to center if needed.

When a worker has finished exploring its instance, only one thread is active.
At this point, the worker first lets center know of its availability by sending an $available$ message (not shown).  After that, it calls the $updateWorkerIPC()$ function in a loop.  This allows the worker process to continue receiving all updates on the best value.  This continues until a $work$ message is received from another process.  The worker lets center know that it started running again, and the received instance can be explored.  
To implement this, the received task could either be taken by the main thread, which will eventually assign tasks to the other threads, or the task could be sent to a thread in a pool, and the current thread would continue looping.

\paragraph{Termination}  
\andres{At the end of each loop}, the center calls a function calls $tryTermination()$ when it detects that all processes are \mlold{idle, which means that they either are in the $AVAILABLE$ state or the $ASSIGNED$ state (see Algorithm~\ref{alg:runcenter}).}  When this happens, a termination signal is sent to every process and they can exit their loop.  \mlold{The reason that center considers a process $w$ as ``not working'' when in the $ASSIGNED$ state is that when the exploration is entirely finished, the process $p$ that is supposed to send a task to $w$ might never do so.  This happens when center assigns $w$ to $p$, and then $p$ finishes exploring just before being notified of this assignment. If $p$ and $w$ are the only processes remaining and all the subtrees have been explored at this point, then $p$ will become $AVAILABLE$ and $w$ will remain $ASSIGNED$, and center should terminate both.}  

However, \mlold{this creates another problem, as we must } check whether every process is truly finished before terminating.  It might occur that every working process is seen as \mlold{available or assigned} by center but that the exploration is not entirely done.  This happens under the following scenario: 1) some process $p$  sends a task $t$ to a process $w$,  \mlold{which is in the ASSIGNED state}; 2) during the transmission of $t$, $p$ \mlold{lets center know that it is available}, and all other workers \mlold{happen to be either available or assigned (including $w$)}; 3) center terminates all processes; 4) the transmission of $t$ to $w$ finishes, but $w$ has been terminated.

We have observed this occurring in practice and devised two safety mechanisms to ensure correct termination:
\begin{enumerate}
    \item 
    workers can keep track of how many tasks they have sent and how many were received.  That is, suppose that $p$ sends a task to $w$ in the \mlold{$updatePendingTasks$ function} of Algorithm~\ref{alg:updateworker}.  The worker increments a variable $nbSentTasks$ by $1$.  
    When $w$ receives, it sends an asynchronous reply to $p$ to acknowledge reception.  Then $p$ receives this and decrements $nbSentTasks$ by $1$.  
    If $p$ receives a termination signal when $nbSentTasks > 0$, it replies $no$ to center and termination is cancelled.  This ensures that termination cannot occur while tasks are being transmitted.

    \item 
    the $tryTermination()$ can have a timeout.  When all workers are seen as available, the function waits for $s$ seconds to see if it receives a $running$ message from a worker.  If it does, termination is cancelled and the loop resumes, and if no such message is received then all processes are sent a termination signal.  If $s$ is large enough, the above scenario can be prevented since $t$ will have time to be transmitted.
    
    Since this timeout will be executed only once (when exploration is truly finished), we can set $s$ large enough without affecting the execution time.  For instance, putting $s = 20$ is inconsequential on the running time of difficult instances that take hours to finish.
\end{enumerate}

Note that the first solution guarantees correct termination, whereas the timeout could still cause problems if communication hangs for more than $s$ seconds.
In practice, the timeout solution by itself with $s = 20$ was verified to work correctly on all instances and is the one currently implemented in GemPBA.

\subsection{Maintaining the most urgent task in workers}
\label{subsec:urgent-task}

We have mentioned several times that the highest priority task should be sent to either free threads or processes, but have not specified how exactly.  
As in~\cite{abu2015scalable}, our point of view is that in a search tree algorithm, the most urgent tasks correspond to highest nodes in the search tree (i.e. those of minimum depth).   
This spreads the exploration across more different parts of the search tree, which allows finding better solutions more quickly, thereby cutting useless branches more quickly as well.
\mlold{Note that although this generic argument applies to most branching algorithms, specific algorithms may benefit from ad hoc task priority functions.  For example, several heuristics explore the most promising child instances first (as in our vertex cover implementation, see Section~\ref{sect:exp}), meaning that the leftmost nodes of the recursion tree could be prioritized.}

\mlold{In any case, for our purposes, }
the recursion tree should be maintained in some way so that at any point, we can access highest nodes when a new task is required to be sent.
Abu-Khzam et al. proposed to assign each node an index based on its location in the tree.  A counter can keep track of the highest priority node and, when it is sent, the counter can be incremented.  This is not too hard to achieve for binary search trees, but this gets more complex for algorithms with higher branching factors, especially when they are heterogeneous across the tree.

Here, we propose an alternate method that is conceptually simple for any branching factor, even if heterogeneous across the search tree.  The idea is simply to store the recursion tree explicitly, in a traditional tree data structure, while ensuring that the size of the tree does not grow exponentially.  
Maintaining a global tree in a multithreaded environment is somewhat complex to do, owing to its dynamic nature.  Instead, we propose that each thread $t_i$ maintains its own task tree $T_i$.  The root of $T_i$ is the task that was initially assigned to $t_i$, with the descending nodes resulting from the recursion.  If needed, the highest priority task can be recovered by inspecting each tree stored by the threads.
As usual, this management should entirely be performed by the library and should be independent of the branching algorithm.

Recall that in Algorithm~\ref{alg:schedalgo}, the $searchTree$ procedure first passes the child instances $I_1, \ldots, I_k$ of parent $I$ to \mlold{$GemPBA.registerChildInstances$}, and then runs $GemPBA.search$ on each subinstance individually.  
The pseudo-code of these routines is illustrated in Algorithm~\ref{alg:dlb}.

\begin{algorithm}
\DontPrintSemicolon
\SetKwProg{Fn}{function}{}{}
\Fn{GemPBA::\mlold{registerChildInstances}($I_1, I_2, \ldots, I_k, parentI$)}
  {
    Let $T_i$ be the task tree of the current thread\;
     \For{$j = 1 .. k$}
     {
        Add $I_j$ as a child of $parentI$ in $T_i$\;
     }
  }
  \Fn{GemPBA::search($I$)}
  {
        \uIf{task $I$ is still in the $T_i$ tree}
        {
            Mark $I$ as ``Exploring'' in $T_i$\;
            searchTree($I$) \tcp*{Explore sequentially}
            Remove task $I$ from $T_i$\;
        
        }

  }     
  \caption{Construction of task tree $T_i$ for thread $t_i$.}
  \label{alg:dlb}
\end{algorithm}
\vspace{3mm}

When an instance $I$ generates child tasks $I_1, \ldots, I_k$ in thread $t_i$, they must be added in the task tree $T_i$\footnote{\mlold{Note that in practice, the instances $I_1, \ldots, I_k$ are copied before being inserted into the task tree. This results in the instances being stored twice: once in the task tree, and once in the recursion tree being explored sequentially.  This could be optimized by maintaining global pointers to tasks, but since task trees have low memory cost, we did not optimize this aspect.}}.  At this point, the update functions in the workers could decide to send one of these tasks at any moment to another thread or process.
This is why the search procedure first checks whether an instance is still present before letting the current thread explore it sequentially.
If there is only one process and one thread, this mimics the sequential version of the search tree exploration.
When a child task and all its descendants are done, it can be removed from the tree.

\paragraph{Size of task trees}
Let us note that the number of nodes in each task tree $T_i$ will always remain proportional to $O(max\_b \cdot D)$, where $max\_b$ is the maximum branching factor and $D$ is the depth of the search in the current thread.  
This is because the topology of the task tree is always a \emph{caterpillar tree}.  That is, each internal node of this tree has at most one child that is another internal node --- the rest are leaf-children.
To see this, it suffices to observe that only tasks of the search tree explored sequentially can have child tasks.  Moreover, when a sequential call is finished, its corresponding task node is removed from the task tree.  It follows that the internal nodes correspond to the path of exploration undertaken by the current thread.
For many branching algorithm, $max\_b$ is a constant and the maximum exploration depth is bounded by the size of the initial instance, in which case the size of each task tree is linear.

\paragraph{Obtaining highest priority tasks}
To find and send the highest priority task, we need a dynamic load balancing (DLB) strategy that sends and removes the first leaf-child of the root of the task tree (the root itself is being explored by the current thread).  
After sending several tasks, it is possible to exhaust all the leaf-children of the root, in which case it has only one child, which is a task on the path currently explored by the current thread.  In this case, the root is of no interest and it can be pruned. Its single child becomes the new root.  \mlold{When a node of the task tree with more than one child is found, we choose its leftmost leaf-child since, as we mentioned, several heuristics prioritize these.  The task corresponding to this child is sent and removed from the tree}.  At this point, it will become the root of the thread or process it is being sent to.
Note that all tree operations can be done in time $O(1)$ with appropriate data structures.

\begin{algorithm}
\DontPrintSemicolon
\SetKwProg{Fn}{function}{}{}
\Fn{GemPBA::sendHighestPriorityTask($dest$)}
  {
     $r = $ root of the task tree $T_i$\;
     $done = False$\;
     \While{not $done$}
     {
        \uIf{$r$ has no children}
        {
            return ``No task"\;
        }
        \uElseIf{$r$ has one child $q$}
        {
            delete $r$\;
            Reroot the tree to $q$, and let $r = q$\;
        }
        \uElse
        {
            Let $\ell$ be \mlold{the leftmost} leaf-child of $r$ not marked as ``Exploring'', \mlold{ and let $I_\ell$ be the instance stored at $\ell$}\;
            Remove $\ell$ from the current tree\;
            $send(\mlold{I_\ell}, dest)$\;
            $done = True$\;
        }
        
     }
  }     
  \caption{Finding a high priority task and updating the task tree.  \mlold{The input $dest$ is either the process or thread to send to. 
  }} 
  \label{alg:sendtask}
\end{algorithm}
\vspace{3mm}

\begin{figure}[H]
    \centering
    \includegraphics[width=0.7\textwidth]{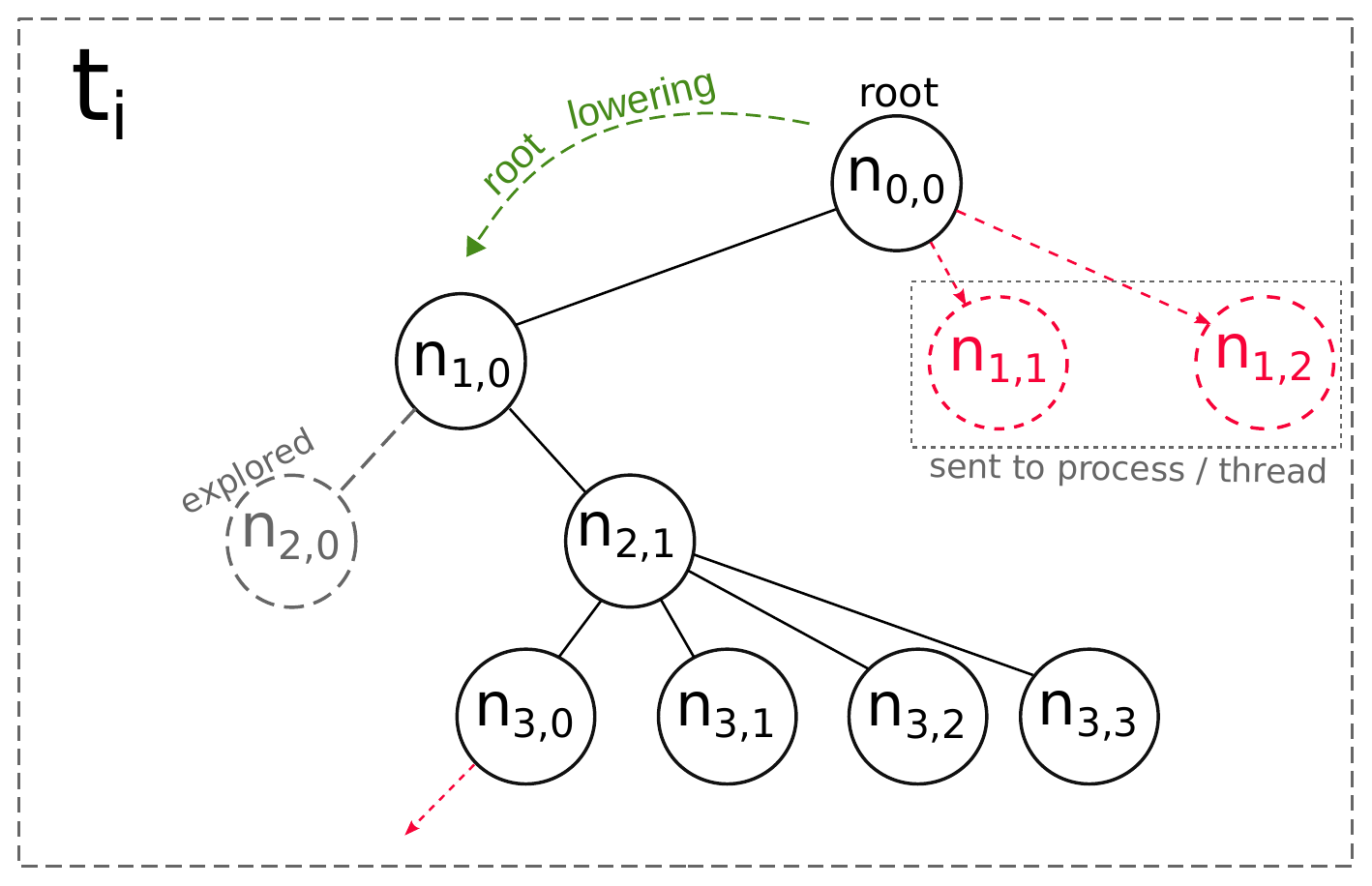}
    \caption{Quasi-horizontal visualisation, heterogeneous search tree.  A node labeled $n_{d, j}$ indicates that its depth in the search tree is $d$, and that it is the $j$-th child of its parent.  Nodes that are dotted have been removed from the tree, either because they were fully explored (gray), or sent to another process/thread (red).
    }
    \label{fig:quasi-horizontal}
\end{figure}

As shown in \cref{fig:quasi-horizontal}, our dynamic load balancing strategy is very simplistic and is not limited to a specific search tree topology. This figure shows the search domain of a single thread $t_i$, where the shown root represents the original task received by the pool thread.

\subsection{Startup phase}

Even though our strategy is fully dynamic and handles arbitrary branching factors, we believe that it should be beneficial to start the exploration in a manner that is as close as possible to the equitable strategy mentioned previously.
Suppose that the branching algorithm never branches into more than $max\_b$ recursive calls.  Here, $max\_b$ may be fixed or may depend on the instance size, but is usually known and can therefore be passed as a parameter to our framework on initialization.  To approximate the equitable strategy, the center first builds a waiting list for every process under the assumption that during the initial exploration of the algorithm, we will most often make $max\_b$ recursive calls.
Under this assumption, 
our aim is that each search tree node at $max\_depth = \log_{max\_b} p$ to be handled by a distinct process.  
It is worth mentioning that in~\cite{abu2015scalable}, the authors also adopted a strategy to achieve equitable startup, albeit in a different manner for their decentralized strategy.

\begin{algorithm}[ht]
    \DontPrintSemicolon
    \SetKwProg{nc}{function}{}{}
    
    \nc{buildWaitingList($p_i$, $base\_d$, $max\_b$, $p$)}
    {   
        //$p_i = $ process index\; 
        //$max\_b = $ maximum branching factor, \;
        //$p $ = number of processes available\;
        //$base\_d$ is the depth of the highest search node assigned to $p_i$\;
        //Initial call is made with $p_i = 1$ and $base\_d = 0$\;
        \For{$d = base\_d .. max\_depth$}
        {
            \For{$j = 1 .. max\_b - 1$}
            {
                $q = (j \times (max\_b)^{d}) + p_{i}$\;
                \uIf{$q \leq p$}
                {
                    Add $q$ to the waiting list of $p_i$\;
                    buildWaitingList($q, d + 1, max\_b, p$)\;
                }
            }
        }
    }
    \caption{Waiting list assignment algorithm.}
    \label{alg:next_child}
\end{algorithm}

\begin{figure}[ht]
    \centering
    \includegraphics[width=0.7\textwidth]{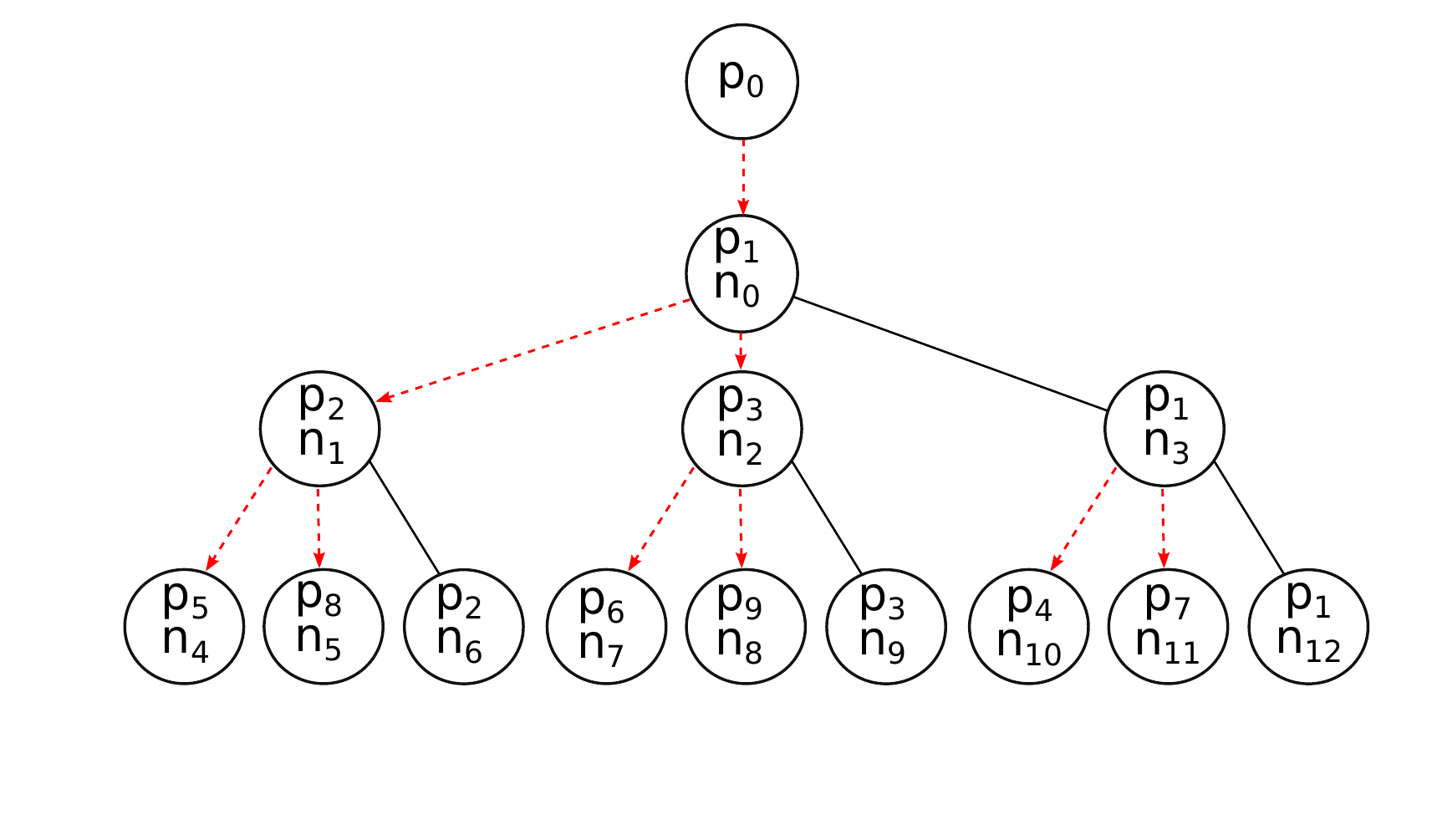}
    \caption{Process topology, where red arrows refer to intended inter-process tasks passing.}
    \label{fig:process_topo}
\end{figure}

\cref{alg:next_child} populates waiting lists to achieve an assignment of search tree nodes.  Importantly, note that even if the branching factors are heterogeneous in the first few recursions, these waiting lists can be used without problem.  This might be slightly suboptimal if the branching factors are far from $max\_b$, but since waiting list items are purged after tasks are sent, this initial assignment only holds at startup and its long term effects should be limited.
The ideal case is illustrated in~\cref{fig:process_topo}, where $max\_b = 3$.
We assume that process $p_i$ will always send its first $max\_b - 1$ search tree tasks to its waiting list, proceed sequentially on the $max\_b-$th task, and repeat the process as it goes deeper.  The waiting list is built accordingly for each process $p_i$, using a parameter $d$ that starts as the depth of the highest search task $p_i$ will be assigned.

In this model, no process is aware of who it is assigned to, and all spawned tasks are sent to the processes in the waiting list in the order of assignment. That is, in~\cref{fig:process_topo}, when sending tasks to other processes,  $p_1$ will send them to $p_2$, $p_3$, $p_4$ and lastly $p_7$ in that order.

\section{Implementation and experimental results}\label{sect:exp}

We have implemented our semi-centralized 
strategy in C++20, using boost~\cite{kormanyos_2018} for task serialization and \emph{OpenMPI 4.0.3} to accomplish inter-process communication.  The implementation should work with any later version.
The code is open-source and available at \href{https://github.com/rapastranac/gempba.git}{https://git.io/Jnx7k}.
Furthermore, we direct the reader to~\cite{pastranamsc} for a more detailed presentation of the implementation.

Using vertex cover as a case study, we first describe how to convert sequential code to parallel concretely, using our implementation.  \mlold{We then describe our implementation of the fully centralized strategy, which was compared against the semi-centralized strategy.  We also describe two graph serialization mechanisms that we evaluated and that have significant impacts on running times.} We then describe the performance results that we obtained on three representative graphs obtained from the Center for Discrete Mathematics and Theoretical Computer Science (DIMACS).  These graphs are routinely used as  challenges to benchmark algorithms~\cite{bader201110th}.  \mlold{Furthermore, we present benchmarks on 100 random graphs.}

\subsection{Converting sequential to parallel}

\noindent
Consider the sequential version of the traditional brute force algorithm for the vertex cover problem, shown in~\cref{alg:mvc_basic}.  Recall that $G - u$ refers to the graph obtained from $G$ after removing $u$ and its incident edges, and $G - N(u)$ is the graph obtained after removing the neighbors of $u$ and their incident edges.
Of course, standard preprocessing rules can be applied on each recursion, for instance removing vertices with zero neighbors.  The concrete rules that we implemented are described in the next subsection.

\begin{algorithm}
\DontPrintSemicolon
\SetKwProg{Fn}{function}{}{}
\Fn{mvc($G, S$)}
  {
    \uIf{$G$ has no edges}
    {
        \uIf{$bestS$ is not initialized or $S.size() < bestS.size()$}
        {
            $bestS = S$\;
        }
        return\;
    }
    /* apply optional preprocessing to $G$ and $S$ */\;
    Find a vertex $u$ of maximum degree in $G$\;
    $G_l = G - u$ and $S_l = S \cup \{u\}$\;
    $G_r = G - N(u)$ and $S_r = S \cup N(u)$\;
    
    $mvc(G_l, S_l)$\;
    $mvc(G_r, S_r)$\;
  }
  
  \caption{Branching algorithm for vertex cover.  The initial call is made with the original graph and $S = \emptyset$.  The $bestS$ variable is global.}
  \label{alg:mvc_basic}
\end{algorithm}
\vspace{3mm}

In order to adapt this code to our framework, the function must include an additional parameter, which is the $gemPBA$ object that handles the parallelism.  This is shown  in~\cref{alg:mvc_gempba}, and the reader will notice that this is simply a special case of~\cref{alg:schedalgo}. Branch calls are delegated to \emph{GemPBA}, whom will decide whether to pass the child calls to a thread or to another process, or to explore them sequentially.

\begin{algorithm}
\DontPrintSemicolon
\SetKwProg{Fn}{function}{}{}
\Fn{mvc($G, S, gemPBA$)}
  {
    $bestS = gemPBA.getBestSolution()$\;
    \uIf{$G$ has no edges}
    {
        \uIf{$S.size() < bestS.size()$}
        {
            $gemPBA.handleSolution(S)$\;
            
        }
        return\;
    }
    /* apply optional preprocessing to $G$ and $S$ */\;
    Find a vertex $u$ of maximum degree in $G$\;
    $G_l = G - u$ and $S_l = S \cup \{u\}$\;
    $G_r = G - N(u)$ and $S_r = S \cup N(u)$\;
    
    \mlold{$I_1 = \{G_l, S_l\}, I_2 = \{G_r, S_r\}, $$parent = \{G, S\}$\;}
    $\mlold{gemPBA.registerChildInstances(I_1, I_2, parent )}$\;
    $gemPBA.search( I_1 )$ \tcp*{Library handles parallelism}
    $gemPBA.search( I_2 )$\;
  }
  
  \caption{The $gemPBA$ version of the vertex cover algorithm.}
  \label{alg:mvc_gempba}
\end{algorithm}
\vspace{3mm}

In our vertex cover implementation, 
we applied the following basic preprocessing rules on $G$ and $S$ in every recursion.  For further details we refer the reader to~\cite{CHEN2001}. 

\begin{itemize}
    \item \textbf{Rule 1:} Remove isolated vertices (i.e. of degree $0$) from $G$.
    \item \textbf{Rule 2:} For each vertex $u$ of degree $1$, add its unique neighbor $v$ to the solution $S$ and remove both $u$ and $v$.
    \item \textbf{Rule 3:} For each vertex $u$ with exactly two neighbors $v$ and $w$ such that $vw$ is an edge, add $v$ and $w$ to the solution and remove $u, v, w$.

\end{itemize}

\noindent
In every recursion, Rules [1-3] are applied iteratively until the graph does not change anymore. 
We used an adjacency matrix bitset implementation to represent our graphs, which allowed for faster union and intersection computations for the reduction rules.

\mlold{
\subsection{Comparison with a centralized scheduler}
We compared our semi-centralized approach to the centralized approach, as described in~\cite{Abu-Khzam2006}.  We  implemented this strategy ourselves mainly because we could not find a centralized implementation on which we could plug in our vertex cover implementation.  Our centralized scheduler uses the same vertex cover code as our semi-central scheduler.  The centralized strategy can be described as follows.

A central process is responsible for receiving tasks and redistributing them to workers (as opposed to workers exchanging tasks between themselves as in our approach).  The center has a limit on how many tasks it can store.  That is, we consider that the center is full if either its memory usage is above a certain configurable threshold, or if it is currently storing more than $c \cdot p$ tasks, where $c$ is a configurable parameter and $p$ is the number of processes.
Each worker stores the status of the center, $full$ or $not full$ (initially set to $not full$).  The center also remembers the state of every worker, which can either be $RUNNING$ or $AVAILABLE$. 
At the startup phase, center passes the original instance to an arbitrary worker.  From this point and onwards, each time a worker registers a new child instance, 
the scheduler checks whether the center is $full$.  If so, the worker adds the task to its local task tree.  If the center is not full, then the worker sends its highest priority task to the center (where priority is defined as in Section~\ref{subsec:urgent-task}).

The center runs in a loop that updates the task queue and the worker states.  
The update function checks whether tasks have been received and adds them to a priority queue if this is the case.  
Then, the update function sends highest prioriy tasks to workers that are in the $AVAILABLE$ state (the chosen workers are updated to the state $RUNNING$).  In our implementation, we used the graph size for priority, with graphs with more nodes having higher priority.
The center then checks whether its task queue is full.  If it was not full on the previous loop and is now full, then center broadcasts a $full$ message to all workers to let them know to stop sending tasks.  If center was full on the previous loop and has become 90\% full, it broadcasts a $not full$ message to workers (we use the 90\% threshold to prevent center from constantly alternating between $full$ and $not full$).
Termination occurs when all workers are in the state $AVAILABLE$ and when the task queue is empty.  Notice that the queue size may go over the desired capacity, since workers could be sending tasks while center is broadcasting its $full$ message.  This is not a problem, since this only results in very few tasks being stored above the limit.

In order to perform a fair comparison with our semi-centralized strategy, we fine-tuned several aspects of the centralized strategy.
A few remarks are noteworthy and possibly of independent interest:

\begin{itemize}
\item we define the center as $full$ if it stores more than $1000p$ tasks or if it takes more than 10 GB memory.  The $1000p$ was determined empirically, and with this threshold the center never reached more than 2GB memory usage.  

\item 
Without a queue size limit, i.e. with only a memory limitation, the center became extremely slow (about 10x slower).  The reason is that the center would never actually reach its memory limit, meaning that workers were \emph{always} sending tasks to center, which induced heavy slowdowns.

\item Using a size-based priority queue for the central tasks list has an important impact on running times.  We tested a first-in first-out queue approach and it was about 2x slower.  Hence, the overhead required to handle the priority queue is worth the effort.  
\end{itemize}

\subsection{Serialization of graphs}

The exact manner in which instances are serialized for inter-process communication can have a significant impact on running times.
Initially, vertex cover instances were encoded as adjacency lists.  More precisely, a graph with $n$ vertices was stored as a list of $n$ lists, with the $i$-th list containing the neighborhood of the $i$-th vertex (encoded as a bitset). 

As it turns out, vertex cover instances can be encoded much more efficiently (albeit with extra programming work), and so we tested a second encoding scheme.  The main observation is that in our vertex cover implementation, each task consists of an induced subgraph of the original instance (i.e. a graph obtained by deleting vertices).  Therefore, at startup each worker is given the original instance, i.e. the full graph.  Then, tasks are encoded with only the set of vertices $X$ that belong to the instance of the graph to be handled.  This set $X$ is encoded with a bit vector of length $n$, where $n$ is the number of vertices in the original graph, and where positions with a $1$ represent vertices in $X$.  Upon receival, a worker can then easily reconstruct the intended instance by taking the subgraph induced by $X$ on the graph it loaded at startup.  As we will see in the experimental results, the overhead introduced by taking induced subgraphs is easily compensated by savings made on communication.  Note however that this approach only works on problems that use induced subgraphs in their exploration.

In what follows, we will call the approach that serializes the whole adjacency lists as the \emph{basic encoding}, and the approach that passes the bit vector representing vertices that are present as the \emph{optimized encoding}.
}

\subsection{Experiments}

We now describe the performance of our approach on DIMACS challenge graphs.

\subsubsection{Experimental setup}
\label{sec:setup}

\mlold{
Computations were performed on three ComputeCanada supercomputers, called  Beluga, Niagara, and Cedar.  The first two provide access to an EDR infiniband (100 Gb/s) network that interconnects all computing nodes, whereas Cedar use Intel Omnipath (v1) interconnections.
Every graph instance was constrained to a maximum running time of 24 hours. Also note that by default, each server has a maximum number of nodes that can be used for one task.
On Niagara, we assigned 20 cores per computing node dedicated to the exploration of instances, up to an allowed maximum of 20 nodes.  On Cedar, we assigned 24 cores per node for a maximum of 16 nodes, and on Beluga, we assigned 20 cores per node for a maximum of 32 nodes.  Note that each core corresponds to a working process, except one core reserved for the center, and that each worker uses two threads, one for exploration and one for communication.  }

We tested the sequential and parallel version of our MVC solver on the following DIMACS graphs.

\begin{itemize}
    \item p\_hat1000-2: 1,000 vertices, 244,799 edges;
    \item p\_hat700-1: 700 vertices, 60,999 edges;
    \item \mlold{DSJ500.5: 500 vertices, 62,624 edges.}
\end{itemize}

\noindent
\mlold{These graphs were chosen because they are representative instances of various levels of difficulty for the vertex cover problem.
The p\_hat1000-2 has average difficulty and takes between one or two days to solve sequentially.  The p\_hat700-1 is a very tough instance that would take, based on our parallel computations, 3-4 weeks to solve sequentially.  The DSJ500.5 graph is easy to solve and takes between one and two minutes to solve sequentially on a standard PC.}
 \mlold{The first two graphs were solved on the Niagara server, DSJ500.5 was solved on the Beluga server, and the random graphs were solved on Cedar (this allowed us to solve instances in parallel).  The harder instances used Niagara since it is less in demand than the others and requires less waiting times.}

\mlold{In order to obtain a larger sample of instances, we also report experiments on random graphs.
To generate these graphs, we specified the number of vertices $n$ and a parameter $p$ between $0$ and $1$. 
 Then, each possible edge between the $n$ vertices was inserted with a probability of $p$ 
(this procedure is known as the $G(n, p)$ Erdos–Rényi model).  The graph is then expected to have $p {n \choose 2}$ edges.  We report the results on 100 graphs of 600 vertices generated with $p = 4/(n - 1)$, so that each vertex is expected to have four neighbors.  A total of 500 real-time hours (i.e. total walltime) was required to solve every instance with every desired number of computing nodes (each instance itself could be solved under 24 hours).  We had to limit the density since higher densities made it too probable to generate instances that could not be solved within 24 hours on one computing node, making it difficult to analyze a large number of instances.  For the same reason, we could not go beyond 600 vertices. }

\subsubsection{Speedups achieved on selected instances}

\begin{figure}[]
        \centering
         \begin{subfigure}[b]{0.495\textwidth}
            \centering
            \vspace*{-0.1cm}
            \includegraphics[width=\textwidth]{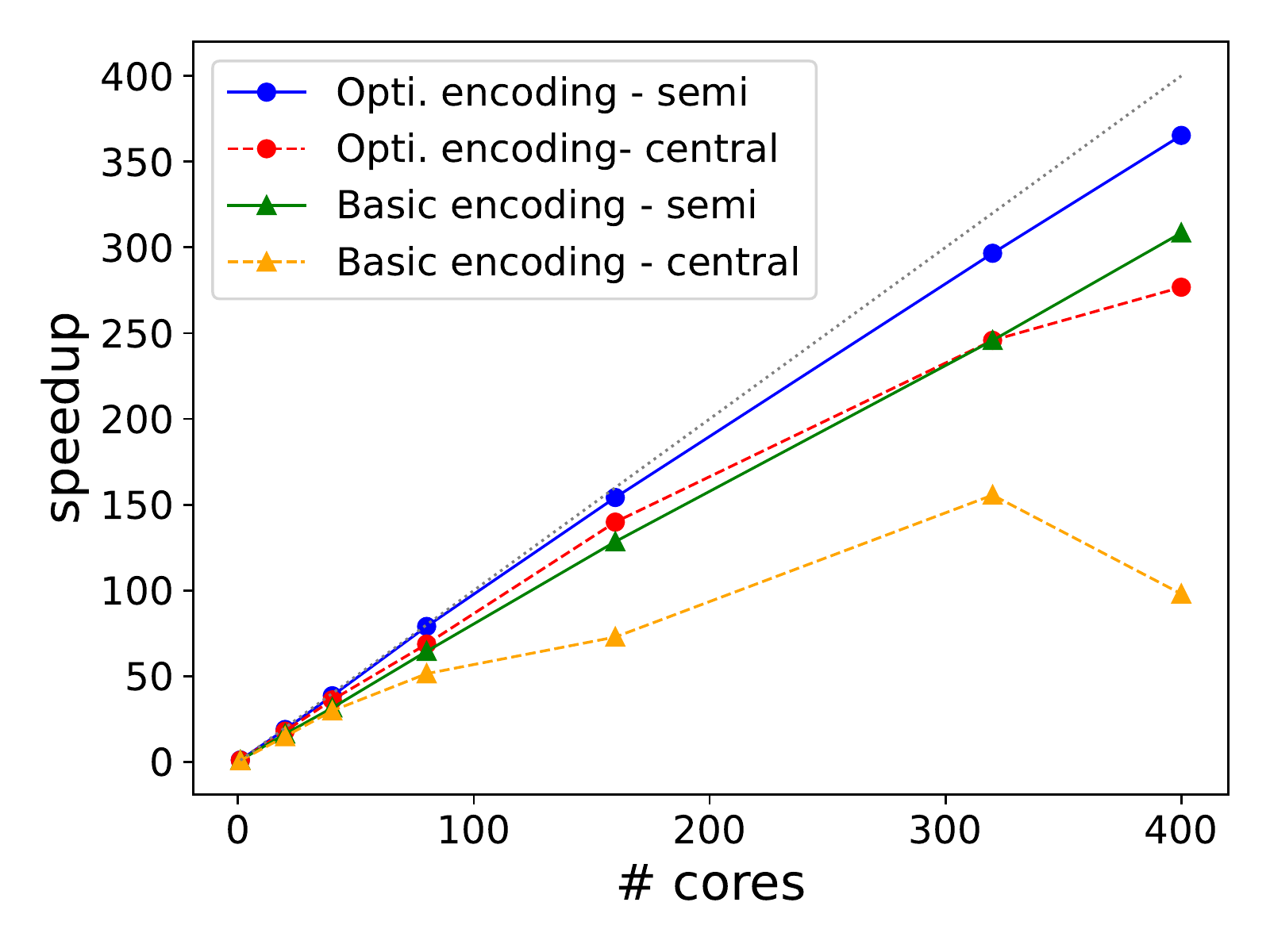}
            \caption[p\_hat1000-2]%
            {{\small p\_hat1000-2}}    
            \label{fig:speedups-phat100}
        \end{subfigure}
        \hfill
        \begin{subfigure}[b]{0.495\textwidth} 
            \centering 
            \vspace*{-0.1cm}
            \includegraphics[width=\textwidth]{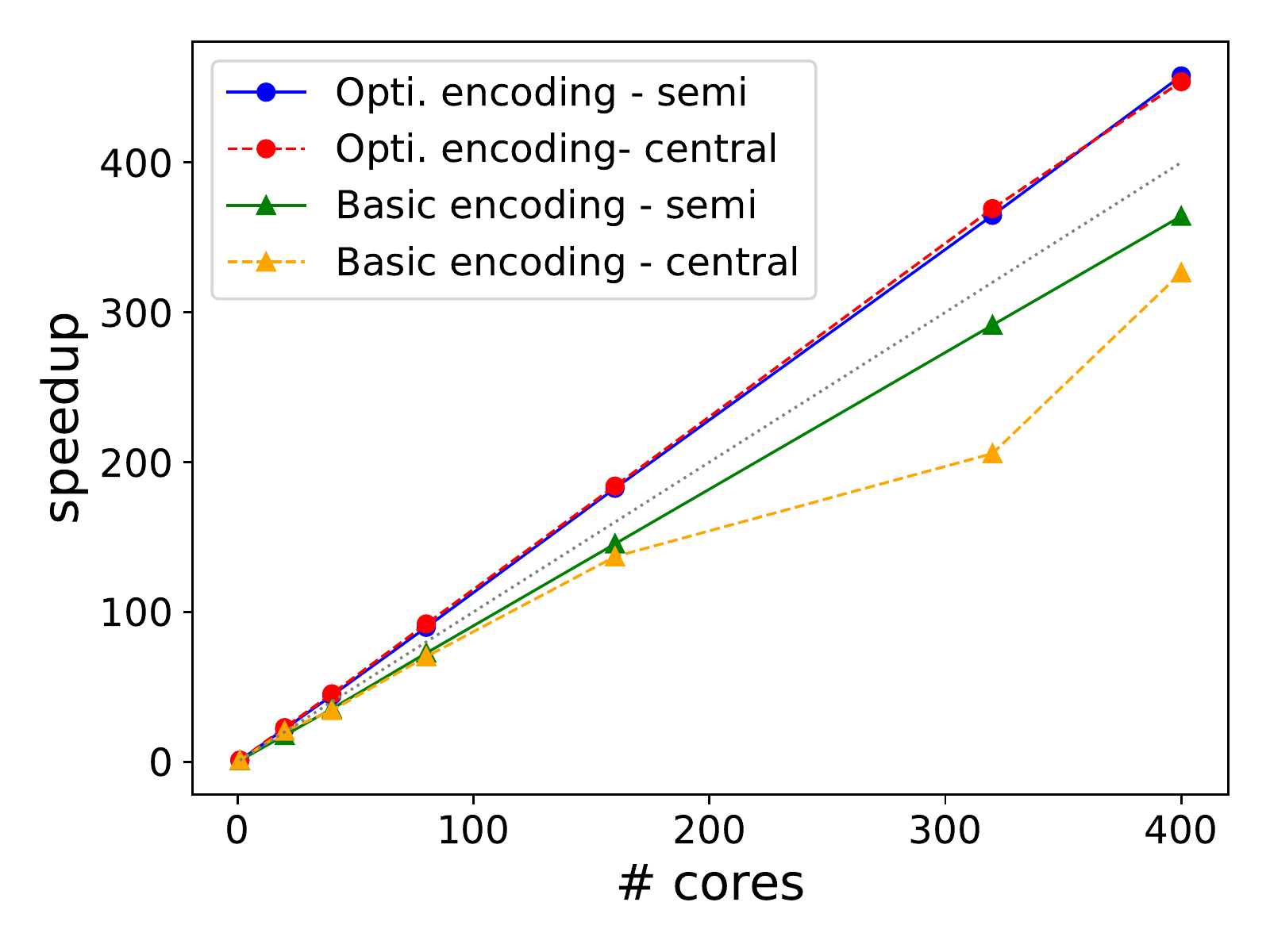}
            \caption[p\_hat700-1]%
            {{\small p\_hat700-1}}    
            \label{fig:speedups-phat700}
        \end{subfigure}
        \begin{subfigure}[b]{0.495\textwidth}
            \centering
            \vspace*{-0.1cm}
            \includegraphics[width=\textwidth]{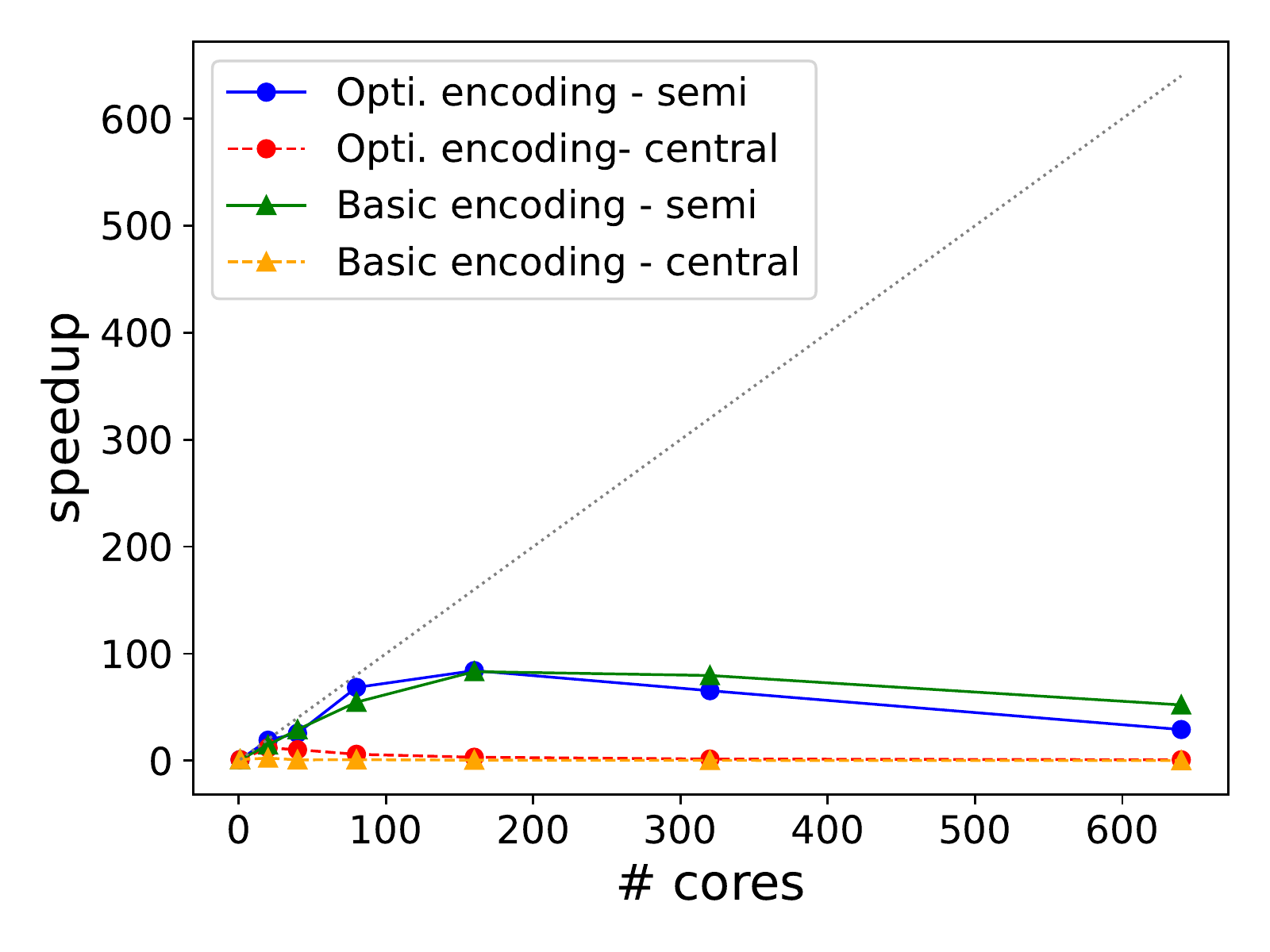}
            \caption[DSJ500.5]%
            {{\small DSJ500.5}}    
            \label{fig:speedups-dsj}
        \end{subfigure}
        \hfill
        \begin{subfigure}[b]{0.495\textwidth}
            \centering
            \vspace*{-0.1cm}
            \includegraphics[width=\textwidth]{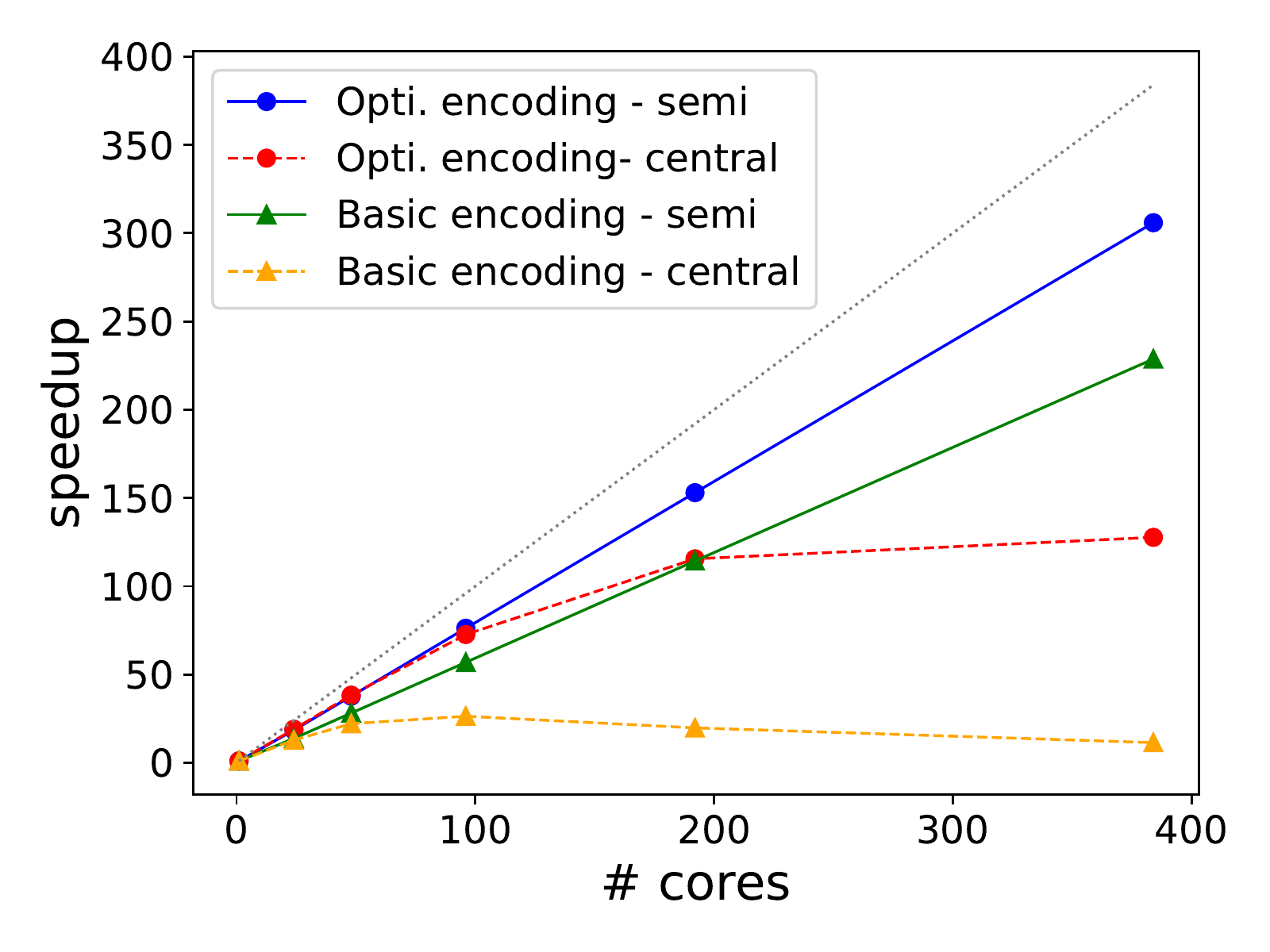}
            \caption[100 random graphs ]%
            {{\small 100 random graphs}}    
            \label{fig:speedups-rnd600}
        \end{subfigure}
        \caption[]
        {Speedups per cores for all instances.}
        \label{fig:speedups}
    \end{figure}

In order to evaluate their scalability, we compared the speedup achieved by each method on the chosen instances, where the methods tested are semi-centralized and centralized with our two possible encoding schemes.  These are illustrated in Figure~\ref{fig:speedups}, which displays speedups achieved for each strategy on the three chosen DIMACS instances and 100 random graphs (the speedup is based on the total time).  The x-axis represents the number of working processes, the y-axis is the speedup with respect to the sequential time.  The grey dotted line is the function $y = x$ and represents an ideal speedup.
The full data that was used to construct these plots can be found in Section~\ref{sec:alltables}.  For each of these graphs, we first assigned one computing node to solve it (with its available cores), then two nodes, and kept doubling the number of nodes until we reached the maximum allowed on each server.   The speedup on $c$ cores is obtained by dividing the sequential running time by the time achieved on $c$ cores.  Note that the sequential times on the two p\_hat graphs were not obtainable within 24 hours, and so it was inferred from the parallel data using exponential regression.  Since every approach gave a different predicted sequential time, we took the average as the canonical sequential time.

\mlold{There are several conclusions that can be drawn from Figure~\ref{fig:speedups}.  
On the top two graphs, we first observe that in terms of speedup, the centralized strategy using the basic encoding underperforms significantly in every case.  The same comment holds for absolute running times, see Table~\ref{tbl:alltimes}.  This can simply be explained by the fact that the centralized strategy needs to communicate instances twice (from a worker to the center, then from the center to a worker), and thus serialization schemes that require more bandwidth incur heavier slowdowns.  The semi-centralized strategy appears to be able to maintain a linear speedup with this encoding. 
The next observation is that using an optimized encoding improves both the centralized and the semi-centralized strategies, with a much more noticeable improvement in the centralized setting.  In fact, with optimized encoding, the centralized strategy is competitive against the semi-centralized strategy, especially on the p\_hat700-1 graph.  This suggests that the type of scheduler does not truly matter and that efforts should rather be invested in encoding tasks efficiently. 
\andres{However, we do observe that the centralized strategy} is less efficient on the random graphs, especially as the number of workers increases.  Whether this tendency persists or not across all instances, we see that the semi-centralized approach is always at least as fast as the centralized approach. 

We have not discussed the DSJ500.5 graph, which is strikingly different owing to its null and even negative slopes on speedups.  As we mentioned, this instance is easy to solve in minutes on a standard computer.  Therefore, assigning up to 640 cores is overkill and results in tremendous computational waste in exchanging tasks.  This confirms the intuition that massive parallelism on exponential algorithms is only truly useful on harder instances, and we leave open the question of determining when it is actually appropriate to deploy all available resources on a given instance.}

\mlold{On random graphs, 
the data is similar to our results on the DIMACS instances.  We see that the running time of the centralized strategy with the basic encoding decreases with the number of nodes, again supporting the idea of heavy computational waste using this approach.   The semi-centralized scheduler appears unaffected by this phenomenon using the basic encoding.  As for the optimized encoding, we see that the running times of both strategies decrease with the assigned number of nodes, as expected.  The centralized strategy is competitive with the semi-centralized until we assigned 16 nodes, where diminishing returns were observed for the centralized idea. 
}

\subsubsection{Detailed data}\label{sec:alltables}

\begin{table}[h]
\setlength\tabcolsep{0pt} 
\begin{tabular*}{\textwidth}{@{\extracolsep{\fill}} *{10}{c} }
\toprule
 \multirowcell{3}{\# nodes} & 
 \multirowcell{3}{\# cores} & 
 \multicolumn{2}{c}{Optimized encoding} & 
 \multicolumn{2}{c}{Large encoding} \\
\cmidrule{3-4} \cmidrule{5-6}
& & \makecell{Semi central} & \makecell{Central} & \makecell{Semi Central} & \makecell{Central} \\

\multicolumn{6}{l}{\textit{p\_hat1000-2}}\\ 
\multicolumn{1}{c}{1}  & \multicolumn{1}{c}{1}  & \multicolumn{1}{c}{\textit{103,968}}  & \multicolumn{1}{c}{\textit{121,140}}  & \multicolumn{1}{c}{\textit{123,012}}  & \textit{143,964}  \\ 
\multicolumn{1}{c}{1}  & \multicolumn{1}{c}{20}  & \multicolumn{1}{c}{6,482}  & \multicolumn{1}{c}{6,845}  & \multicolumn{1}{c}{7,545}  & 8,279  \\ 
\multicolumn{1}{c}{2}  & \multicolumn{1}{c}{40}  & \multicolumn{1}{c}{3,185}  & \multicolumn{1}{c}{3,397}  & \multicolumn{1}{c}{3,896}  & 4,121  \\ 
\multicolumn{1}{c}{4}  & \multicolumn{1}{c}{80}  & \multicolumn{1}{c}{1,557}  & \multicolumn{1}{c}{1,789}  & \multicolumn{1}{c}{1,907}  & 2,389  \\ 
\multicolumn{1}{c}{8}  & \multicolumn{1}{c}{160}  & \multicolumn{1}{c}{798}  & \multicolumn{1}{c}{880}  & \multicolumn{1}{c}{958}  & 1,689  \\ 
\multicolumn{1}{c}{16}  & \multicolumn{1}{c}{320}  & \multicolumn{1}{c}{415}  & \multicolumn{1}{c}{500}  & \multicolumn{1}{c}{501}  & 791  \\ 
\multicolumn{1}{c}{20}  & \multicolumn{1}{c}{400}  & \multicolumn{1}{c}{337}  & \multicolumn{1}{c}{444}  & \multicolumn{1}{c}{399}  & 1,255  \\ 
\gape{}\\
\multicolumn{6}{l}{\textit{p\_hat700-1}}    \\ 
\multicolumn{1}{c}{1}  & \multicolumn{1}{c}{1}  & \multicolumn{1}{c}{\textit{2,015,172}} & \multicolumn{1}{c}{\textit{1,869,300}} & \multicolumn{1}{c}{\textit{2,478,060}} & \textit{2,120,832} \\ 
\multicolumn{1}{c}{1}  & \multicolumn{1}{c}{20}  & \multicolumn{1}{c}{\textit{96,516}}  & \multicolumn{1}{c}{\textit{92,952}}  & \multicolumn{1}{c}{\textit{119,880}}  & \textit{103,104}  \\ 
\multicolumn{1}{c}{2}  & \multicolumn{1}{c}{40}  & \multicolumn{1}{c}{47,942}  & \multicolumn{1}{c}{46,742}  & \multicolumn{1}{c}{59,985}  & 61,744  \\ 
\multicolumn{1}{c}{4}  & \multicolumn{1}{c}{80}  & \multicolumn{1}{c}{23,636}  & \multicolumn{1}{c}{23,040}  & \multicolumn{1}{c}{29,225}  & 30,215  \\ 
\multicolumn{1}{c}{8}  & \multicolumn{1}{c}{160}  & \multicolumn{1}{c}{11,618}  & \multicolumn{1}{c}{11,525}  & \multicolumn{1}{c}{14,582}  & 15,488  \\ 
\multicolumn{1}{c}{16}  & \multicolumn{1}{c}{320}  & \multicolumn{1}{c}{5,815}  & \multicolumn{1}{c}{5,744}  & \multicolumn{1}{c}{7,277}  & 10,306  \\ 
\multicolumn{1}{c}{20}  & \multicolumn{1}{c}{400}  & \multicolumn{1}{c}{4,633}  & \multicolumn{1}{c}{4,672}  & \multicolumn{1}{c}{5,824}  & 6,496  \\ 
\gape{}\\
\multicolumn{6}{l}{\textit{DSJ500.5}}    \\ 
\multicolumn{1}{c}{1}  & \multicolumn{1}{c}{1}  & \multicolumn{1}{c}{73.24}  & \multicolumn{1}{c}{73.24}  & \multicolumn{1}{c}{73.24}  & 73.24  \\ 
\multicolumn{1}{c}{1}  & \multicolumn{1}{c}{20}  & \multicolumn{1}{c}{3.83}  & \multicolumn{1}{c}{6.15}  & \multicolumn{1}{c}{5}  & 30.21  \\ 
\multicolumn{1}{c}{2}  & \multicolumn{1}{c}{40}  & \multicolumn{1}{c}{2.86}  & \multicolumn{1}{c}{7.13}  & \multicolumn{1}{c}{2.52}  & 98.24  \\ 
\multicolumn{1}{c}{4}  & \multicolumn{1}{c}{80}  & \multicolumn{1}{c}{1.07}  & \multicolumn{1}{c}{12.28}  & \multicolumn{1}{c}{1.34}  & 81.53  \\ 
\multicolumn{1}{c}{8}  & \multicolumn{1}{c}{160}  & \multicolumn{1}{c}{0.87}  & \multicolumn{1}{c}{22.82}  & \multicolumn{1}{c}{0.88}  & 148.92  \\ 
\multicolumn{1}{c}{16}  & \multicolumn{1}{c}{320}  & \multicolumn{1}{c}{1.12}  & \multicolumn{1}{c}{47.53}  & \multicolumn{1}{c}{0.92}  & 297.57 \\
\multicolumn{1}{c}{32}  & \multicolumn{1}{c}{640}  & \multicolumn{1}{c}{2.55}  & \multicolumn{1}{c}{86.90}  & \multicolumn{1}{c}{1.46}  & 579.59 \\
\gape{}\\
\multicolumn{6}{l}{\textit{100 random graphs (total running time)}}   \\ 
\multicolumn{1}{c}{1}  & \multicolumn{1}{c}{1}  & \multicolumn{1}{c}{1,090,584}  & \multicolumn{1}{c}{1,090,584}  & \multicolumn{1}{c}{1,090,584}  & 1,090,584  \\ 
\multicolumn{1}{c}{1}  & \multicolumn{1}{c}{24}  & \multicolumn{1}{c}{59,074}  & \multicolumn{1}{c}{57,405}  & \multicolumn{1}{c}{79,448}  & 85,002  \\ 
\multicolumn{1}{c}{2}  & \multicolumn{1}{c}{48}  & \multicolumn{1}{c}{28,937}  & \multicolumn{1}{c}{28,485}  & \multicolumn{1}{c}{38,792}  & 49,330  \\ 
\multicolumn{1}{c}{4}  & \multicolumn{1}{c}{96}  & \multicolumn{1}{c}{14,315}  & \multicolumn{1}{c}{15,016}  & \multicolumn{1}{c}{19,179}  & 41,421  \\ 
\multicolumn{1}{c}{8}  & \multicolumn{1}{c}{192}  & \multicolumn{1}{c}{7,129}  & \multicolumn{1}{c}{9,442}  & \multicolumn{1}{c}{9,542}  & 55,242  \\ 
\multicolumn{1}{c}{16}  & \multicolumn{1}{c}{384}  & \multicolumn{1}{c}{3,565}  & \multicolumn{1}{c}{8,542}  & \multicolumn{1}{c}{4,769}  & 95,811  \\ 


\midrule 
\bottomrule
\end{tabular*}
\caption{Running times in seconds for each chosen graph.  
Values in italics were inferred using exponential decay regression (first row of p\_hat1000-2, first two rows of p\_hat700-1). 
Since sequential running times were unavailable for these graphs, we used the average of these inferred values as the sequential time to calculate the speedups presented in Figure~\ref{fig:speedups}.}
\label{tbl:alltimes}
\end{table}

\mlold{
Table~\ref{tbl:alltimes} 
displays detailed information on the experiments, which were used to create the plots shown above.  The two p\_hat graphs and the DSJ500.5 graphs were run on Niagara and Beluga, respectively, with 20 cores per node allowing up to 20 and 16 nodes, respectively, and the 100 random graphs on Cedar with 24 cores per node for up to 16 nodes.
The first row of each table shows the sequential running times, which we recall had to be inferred using exponential regression.  Since sequential times were required to estimate speedups, we used the average of these inferred values to compute the speedup.
}

\section{Conclusion and future work}

\mlold{In this work, we have devised a novel load balancing strategy for the massive parallelization of recursive branching algorithms.  It is semi-centralized, in the sense that it borrows ideas from both centralized and decentralized schemes.  Our strategy does make use of a central coordination process, but task exchanges are delegated to working processes as in a decentralized setting.  By minimizing the center responsibilities and guaranteeing successful task requests, we limit the possibility of bottlenecks at the center.  

In our comparative analysis with a fully centralized scheduler, we saw that it never outperformed the semi-centralized scheduler.  In fact, in order to make the centralized strategy competitive, we had to devote efforts to devising an optimized serialization scheme --- which can be spared by using our approach.  

There is still much to explore in this line of research.  We used the classical vertex cover algorithm as a proof-of-concept, but it will be necessary to test other instances and even other branching algorithms with different branching factors to fully validate our strategy.  \ml{In particular, the PACE challenge of 2019 consisted in finding exact solutions to a set of vertex cover instances, which we may consider in future work (see~\cite{hespe2020wegotyoucovered}). 
 Moreover, challenges from other years include problems that can be also solved using branching algorithms.}  It also remains to compare our strategy with a fully decentralized strategy
\andres{(unfortunately, it seems that this is not implemented in open source software, and implementing it is not a trivial task).} 

\ml{The strategies for storing high-priority tasks  could also be explored.  In~\cite{abu2015scalable}, the authors propose a numbering scheme to represent the tree of tasks, thereby avoiding to store the whole task tree in memory.  Although allowing such a scheme in a generic framework will be technically difficult, it would be interesting to evaluate how much this type of lighter strategy impacts memory and running times.
Also, as we showed that the size of the serialized instances heavily impacts running times, other ideas may be explored.  For instance, in vertex cover we could encode an instance by only storing the list of vertices that have been included in the solution so far, which is enough to infer the current instance and should be smaller.  More generally, the encoding could consist of the task tree nodes indices that lead to the current instance (e.g. using a numbering scheme), which are likely to be small and applicable to any problem.  
}

Finally, we should mention that a more detailed analysis of the distribution of task requests over time would be interesting.  Indeed, we believe that on difficult instances, the workers perform few requests since they have large instances to explore, whereas near the end of the execution, only small tasks remain.  
\andres{Based on this observation, we can hypothesize that most requests occur during this phase and that they contribute the most to the overall running time.}   \ml{One solution could be to let the workers solve instances sequentially once the instance size is below a certain threshold.}
More experiments are required to validate this hypothesis and measure its impact on running time --- after which further research can focus on this problem.  We also note that our second serialization scheme, which has the best performance, always requires $n$ bits, in every exchange.  There could be a point in the execution where the nodes could coordinate and eliminate vertices that need no consideration, in order to reduce $n$ during the exploration.  This might especially help in the later stages of the execution.
}

\section*{Acknowledgements}

\ml{The authors sincerely thank the anonymous reviewers for their very useful comments on improving the quality of the paper, and on their insightful suggestions for future work.}

Manuel Lafond acknowledges financial support for this project from the Natural Sciences and Engineering Research Council of Canada (NSERC) and from the Fonds de recherche du Québec Nature and technologies (FRQNT).
Andres Pastrana would like to thank to Dr. Daniel Gruner and the Scinet team for providing access to Niagara Supercomputer.
The authors would like to thank Compute Canada for providing access to the other clusters during the development of this project.

\vspace{-5mm}

 \bibliographystyle{unsrtnat} 
 \bibliography{cas-refs}


\newpage

\appendix
\setcounter{page}{1}

\section{Flowcharts}
\label{sec:sample:appendix}

\begin{figure}[H]
    \centering
    \includegraphics[width=0.9\textwidth]{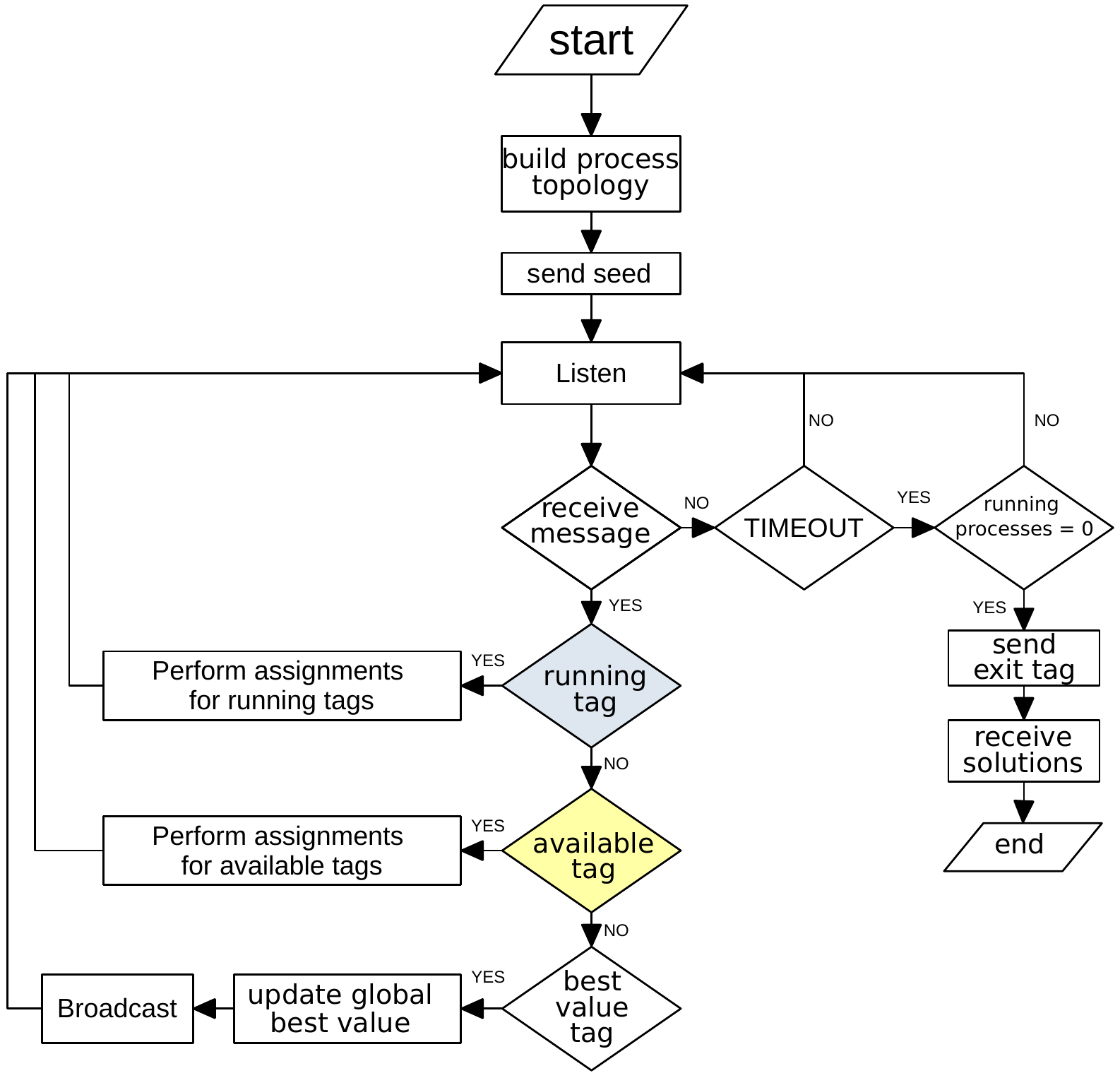}
    \caption{Center process flowchart}
    \label{fig:center}
\end{figure}

\begin{figure}[b]
    \centering
    \includegraphics[width=0.7\textwidth]{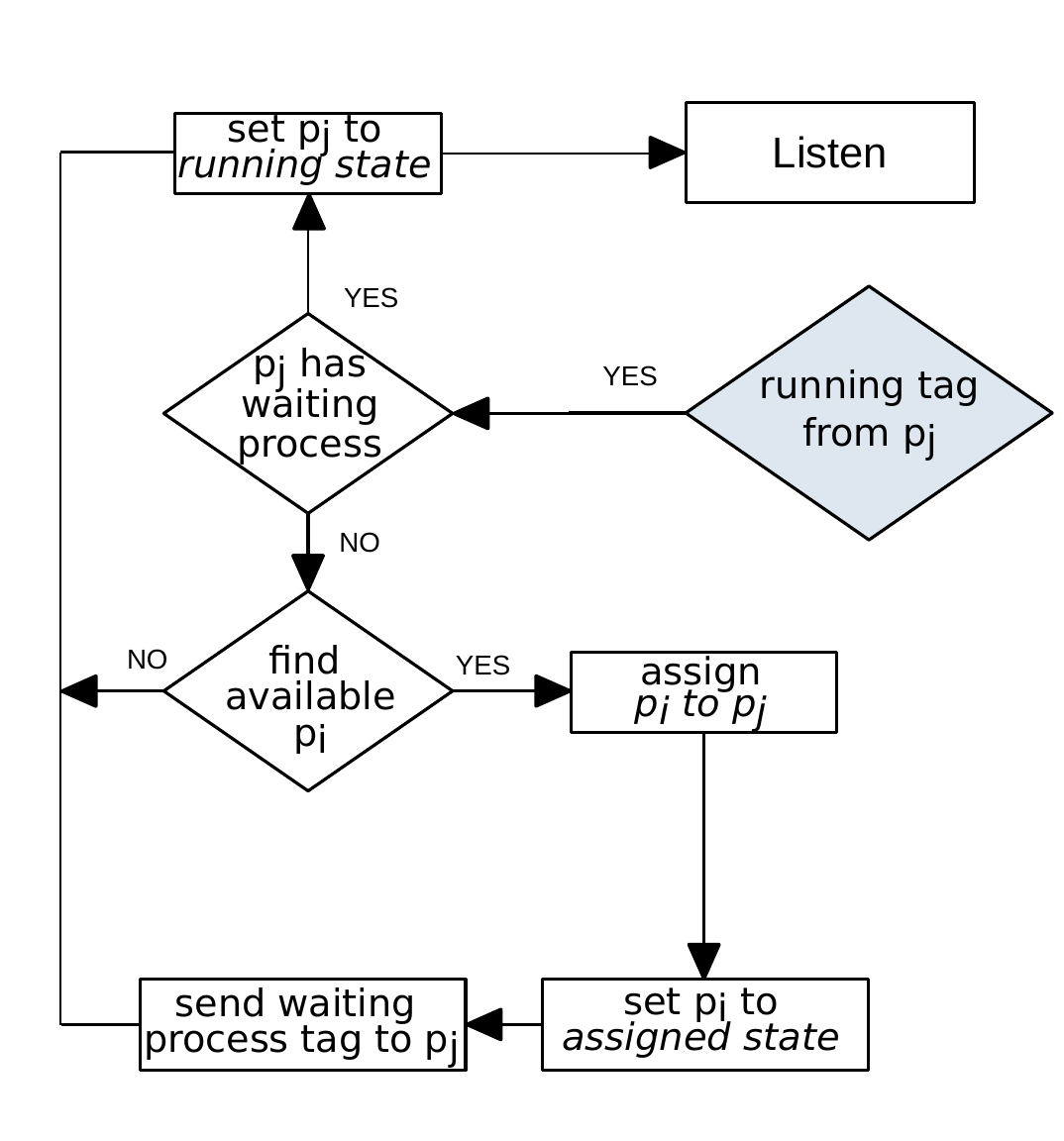}
    \caption{Perform assignments for running tags.}
    \label{fig:run_tag}
\end{figure}

\begin{figure}
    \centering
    \includegraphics[width=0.6\textwidth]{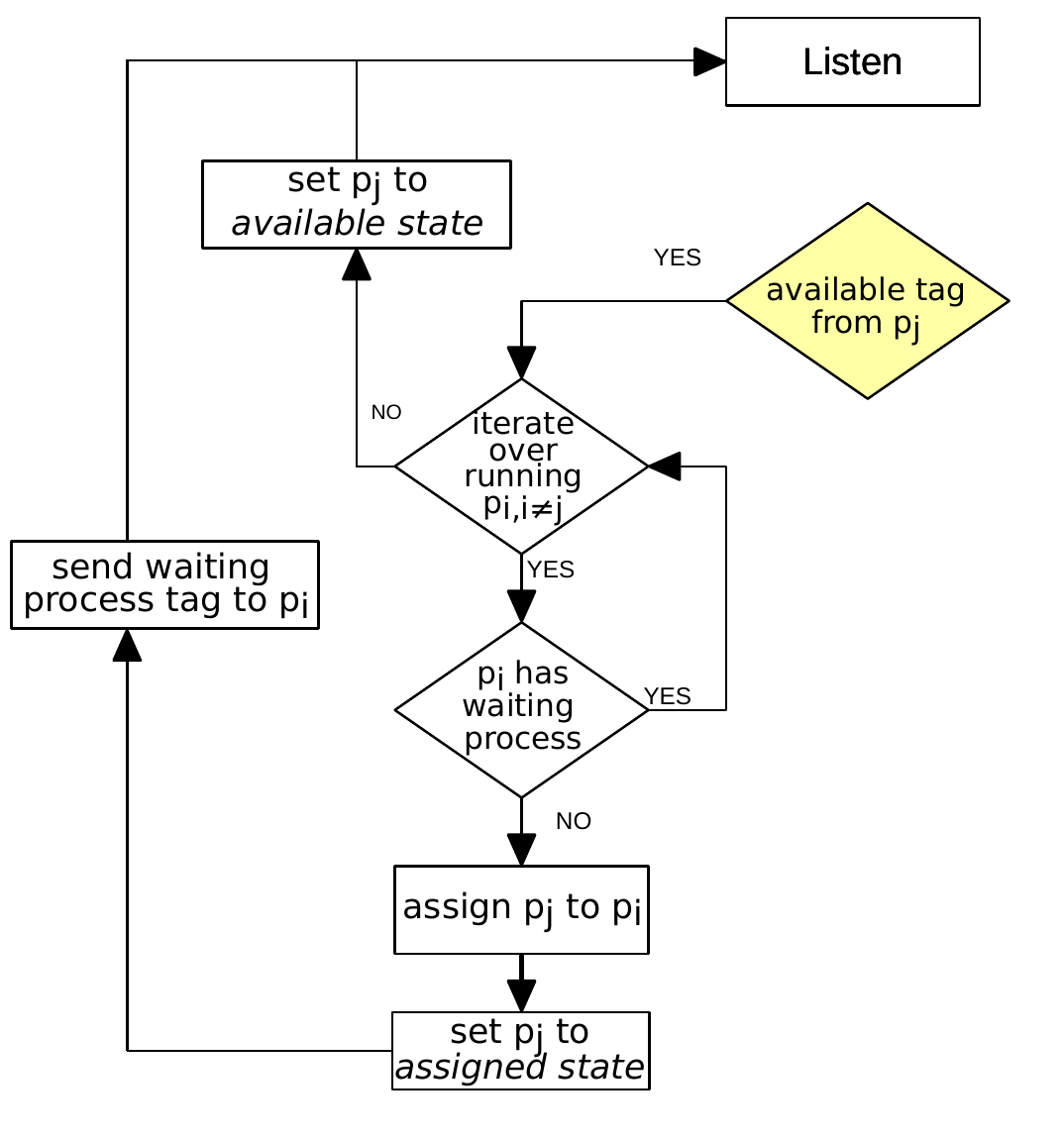}
    \caption{Perform assignments for available tags.}
    \label{fig:avail_tag}
\end{figure}

\begin{figure}
    \centering
    \includegraphics[width=0.6\textwidth]{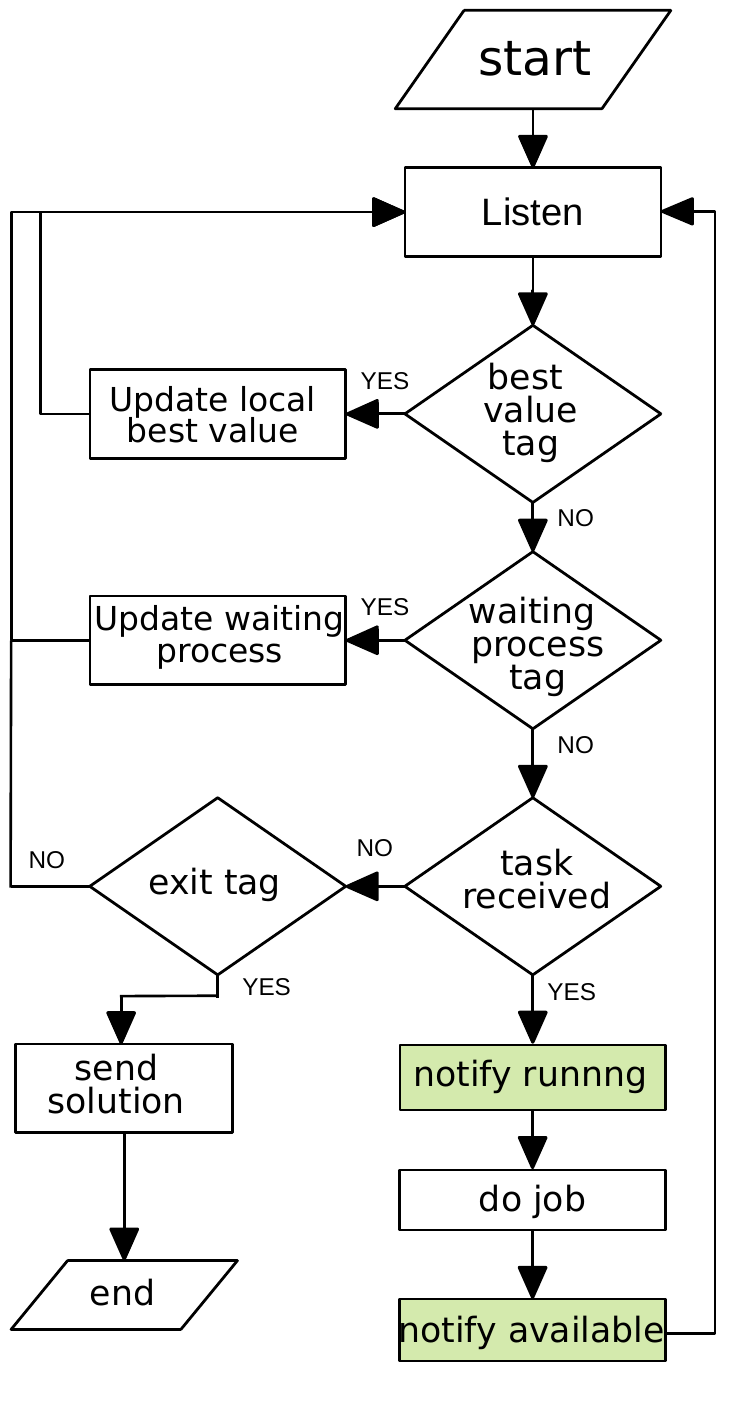}
    \caption{Worker process}
    \label{fig:worker}
\end{figure}

\begin{figure}
    \centering
    \includegraphics[width=0.6\textwidth]{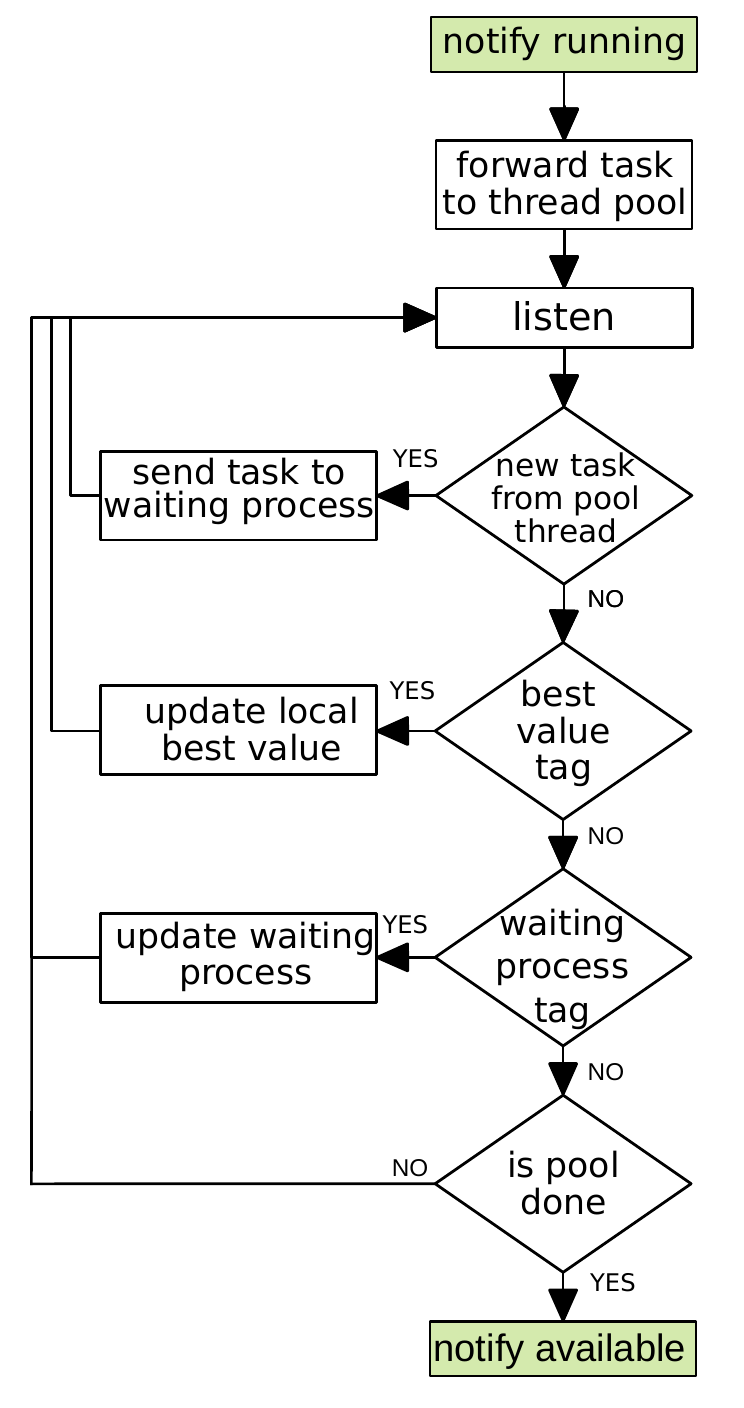}
    \caption{Worker process, do job details}
    \label{fig:worker2}
\end{figure}

In this section we present a more detailed flowchart of our implementation for this study, where \cref{fig:center} shows the series of decisions made by the center process. \cref{fig:run_tag,fig:avail_tag} are the expansion of the description boxes when the center process receives \emph{running} and \emph{available} notifications as \emph{tags}. In~\cref{fig:worker}, we show the decisions by the worker process, $p_i$, which is in constant listening mode from any other process and start working when it receives a message containing a task. The series of decisions made when it is doing its job, is expanded in~\cref{fig:worker2}, where the main thread $t_0$ of this process is in charge of the inter-process communication. Not shown here is the pool threads, $t_i,\dots,t_{c-1} $, which receives the initial task from the main thread, then it is partitioned into other tasks and sent accordingly to waiting processes or waiting threads availability, prioritizing processes by default.

\end{document}